\documentclass[12pt]{article} 
\usepackage[sectionbib]{natbib}
\usepackage{array,epsfig,fancyheadings,rotating}
\usepackage[]{hyperref}  
\usepackage{sectsty, secdot}
\usepackage{color}
\sectionfont{\fontsize{12}{14pt plus.8pt minus .6pt}\selectfont}
\renewcommand{\theequation}{\thesection\arabic{equation}}
\subsectionfont{\fontsize{12}{14pt plus.8pt minus .6pt}\selectfont}

\newcommand{\bG}{{\bf G}}
\newcommand{\bS}{{\bf S}}

\newcommand{\bX}{{\bf X}}

\newcommand{\bh}{{\bf h}}
\newcommand{\bb}{{\bf b}}
\newcommand{\bu}{{\bf u}}
\newcommand{\bo}{{\bf o}}
\newcommand{\bs}{{\bf s}}
\newcommand{\bd}{{\bf d}}
\newcommand{\bbeta}{\boldsymbol{\beta}}

\newcommand{\bgamma}{\boldsymbol{\gamma}}

\newcommand{\btheta}{\boldsymbol{\theta}}

\newcommand{\bupsilon}{\boldsymbol{\upsilon}}

\newcommand{\bSigma}{\boldsymbol{\Sigma}}

\newcommand{\T}{\!\top\!}
\newcommand{\trans}{^{\mbox{\begin{tiny}{\sf T}\end{tiny}}}}

\usepackage{amsmath}
\usepackage{amssymb}
\usepackage{amsfonts}
\usepackage{multirow}
\usepackage{amsthm}

\allowdisplaybreaks

\newtheorem{theorem}{Theorem}
\newtheorem{lemma}{Lemma}
\newtheorem{corollary}{Corollary}

\theoremstyle{definition}

\begin{document}


\renewcommand{\baselinestretch}{1}

%
%


\fontsize{12}{14pt plus.8pt minus .6pt}\selectfont \vspace{0.8pc}
\centerline{\large\bf Quantile Regression Modeling of Recurrent Event Risk}
\vspace{2pt} \centerline{Huijuan Ma$^1$, Limin Peng$^2$, Chiung-Yu Huang$^3$ and Haoda Fu$^4$} \vspace{2pt}
\centerline{\it $^{1}$Institute of Statistics and Interdisciplinary Sciences, East China Normal University}  \vspace{2pt}
\centerline{\it $^{2}$Department of Biostatistics and Bioinformatics, Emory University}  \vspace{2pt}
\centerline{\it $^{3}$Department of Epidemiology and Biostatistics, University of California, San Francisco}  \vspace{2pt}
\centerline{\it $^{4}$Eli Lilly and Company}
\vspace{.35cm} \fontsize{9}{11.5pt plus.8pt minus .6pt}\selectfont


\noindent {\it Abstract:}
Progression of chronic disease is often manifested by repeated occurrences of disease-related events over time. Delineating the heterogeneity in the risk of such recurrent events can provide valuable scientific insight for guiding customized disease management. In this paper, we present a new modeling framework for recurrent event data, which renders a flexible and robust characterization of individual multiplicative risk of recurrent event  through quantile regression that accommodates both observed covariates and unobservable frailty.   The proposed modeling requires no distributional specification of the unobservable frailty, while permitting the exploration of dynamic covariate effects. We develop estimation and inference procedures for the proposed model through a novel adaptation of the principle of conditional score. The asymptotic properties of the proposed estimator, including the uniform consistency and weak convergence, are established. Extensive simulation studies demonstrate satisfactory finite-sample performance of the proposed method. We illustrate the practical utility of the new method via an application to a diabetes clinical trial that explores the risk patterns of hypoglycemia in Type 2 diabetes patients.

\vspace{9pt}
\par

\def\thefigure{\arabic{figure}}
\def\thetable{\arabic{table}}

\renewcommand{\theequation}{\thesection.\arabic{equation}}

\fontsize{12}{14pt plus.8pt minus .6pt}\selectfont

\setcounter{equation}{0} 
\section{Introduction}

In chronic disease follow-up studies, recurrent events, such as tumor recurrence or repeated  adverse events, are often captured over time to track the progression of the disease. Understanding how baseline characteristics influence the risk of recurrent events can provide useful information to guide disease management. To this end, various methods have been developed in literature, including regression modeling of the multivariate time to recurrent events \citep[among others]{Wei1989, Cai1995, Spiekerman1998}, the gap time between recurrent events \citep[among others]{Wang1999,  Chang1999, Pena2001, Luo2013}, and the counting process of recurrent events based on the intensity function \citep[among others.]{Prentice1981, Andersen1982} or the marginal mean/rate functions \citep[among others]{Pepe1993, Lawless1995, Lin2000, Schaubel2006}.  Readers may refer to \cite{Cook2007}  for a comprehensive review of statistical methods for recurrent event data.

In this work, we develop a new counting-process based approach, which 
offers enhanced flexibility and robustness for delineating the heterogeneity in recurrent event risk.
The foremost step of our proposal is to sensibly quantify subject-specific risk of recurrent events through a novel view of a multiplicative  intensity model. More specifically, let $N_i^*(t)$ denote the underlying counting process of recurrent events (i.e. the number of recurrent events that have occurred by time $t$). Following \cite{Wang2001}, we model the occurrence of recurrent events by  a subject-specific nonstationary Poisson process. That is, we assume that $N_i^*(t)$, given a nonnegative random variable $\gamma_i$, is a nonstationary Poisson process with the intensity function,
\begin{equation}\label{model1}
\lambda(t|\gamma_i)=\gamma_i\cdot \lambda_0(t),
\end{equation}
where the baseline intensity $\lambda_0(t)$ is an unknown, nonnegative, and  continuous function.

Model \eqref{model1} represents a general formulation of the multiplicative intensity model. Here $\gamma_i$ captures the scale shift of  subject $i$'s intensity process from the baseline intensity, and can be interpreted as the latent subject-specific risk of recurrent events. As a subtle but important distinction from the modeling framework of \cite{Wang2001} and others \citep[e.g.]{Nielsen1992, Oakes1992}, we shall utilize $\gamma_i$ to account for the individual differences (in recurrent event occurrences) explained by either the observed covariates $\tilde\bX_i$ or   some unobservable, nonnegative frailty $\xi_i$. In the existing work, a frailty $\xi_i$ was typically used to accommodate the within-subject dependency of recurrent events, and the influence of $\tilde\bX_i$  was often addressed separately.
It is easy to see that the proportional intensity model \citep{Andersen1982} is a special case of model \eqref{model1} with $\gamma_i=\exp(\tilde\bX_i\trans\bb_0)$. When $\gamma_i=\xi_i\exp(\tilde\bX_i\trans\bb_0)$, model \eqref{model1} reduces to  \cite{Wang2001}'s semiparametric multiplicative intensity model.

The general representation of the multiplicative intensity model in \eqref{model1} enlightens a new modeling perspective for addressing the targeted question of how the recurrent event risk during follow-up is influenced by baseline covariates. 
Our basic idea is to view $\gamma_i$ as a latent individual risk measure and link $\log\gamma_i$ with $\tilde\bX_i$ through a linear quantile regression model \citep{Koenker1978}. This effort would lead to a broader class of multiplicative intensity models, because existing ones essentially exert a linear model for  $\log\gamma_i$. For example,  \cite{Wang2001}'s model is equivalent to specifying $\log\gamma_i=\log\xi_i+\tilde\bX_i\bb_0$, which is a special case of a linear quantile regression model.

There are multi-fold practical benefits from modeling $\log\gamma_i$ by linear quantile regression. First, a dynamic relationship between $\tilde\bX_i$ and $\gamma_i$ is often scientifically important. For example, a test treatment may be more beneficial for subjects prone to a high risk of recurrent events, compared to subjects of a low risk of recurrent events. Adopting a quantile regression model  allows us to flexibly explore  such dynamic associations of interest, which, however, are not permitted by a linear model for $\log\gamma_i$. Secondly,  classic multiplicative intensity models may be preferred in practice for reasons such as interpretation simplicity.  Inference tools developed for the quantile regression of  $\gamma_i$ can be used to test the goodness-of-fit of these models and thus ensure the validity of the results. Of note, there are other applications or generalizations of quantile regression to recurrent event data.
For example, \cite{Luo2013} studied the quantile regression modeling of gap times between recurrent events. \cite{Huang2009} and \cite{Sun2016} proposed  the accelerated recurrence time model, which reduces to a quantile regression model when the event of interest is not recurrent.  Compared to these existing methods, the new modeling framework  proposed in this work  directly applies the quantile regression strategy to a sensible measure of subject-specific recurrent event risk; thus
can yield more straightforward interpretations regarding the heterogeneity in recurrent event risk that is of interest.

To tackle the quantile regression problem for the recurrent event risk captured by $\gamma_i$, a main technical difficulty relates to the latent nature of $\gamma_i$. We address this challenge by employing the principle of conditional score in the settings of quantile regression and multiplicative intensity regression of recurrent events. It is worth emphasizing that the quantile regression modeling of $\gamma_i$ can naturally account for 
an unobservable frailty $\xi_i$ without requiring distributional assumptions. Revising the conditional likelihood arguments in \cite{Wang2001}, we can also circumvent any parametric specification of the baseline intensity function $\lambda_0(t)$ in model \eqref{model1}. These nice properties indicate the flexibility and robustness of the proposed regression framework.

In the rest of the paper,  we further elaborate the proposed modeling in Section 2. 
In Section 3, we present the proposed estimation and inference procedures as well as the asymptotic studies. We report our numerical studies in Section 4, including results from Monte-Carlo simulations, and an application to the DURABLE study, a randomized clinical trial in Type-2 diabetes patients.  A few remarks are concluded in Section 5.

\section{The Proposed Model}

We begin with the introduction of  data and notation.  Let $T_i^{(j)}$ denote the time to the $j$th recurrent event  of subject $i$. The underlying counting process for the recurrent events is defined as $N_i^*(t) = \sum_{j=1}^\infty I( T_i^{(j)} \leq t )$. Suppose the observation of recurrent events is terminated by a censoring time $C_i$. The observed counting process is then given by $N_i(t)=N_i^*(t \wedge C_i) = \sum_{j=1}^\infty I( T_i^{(j)} \leq t \wedge C_i)$, where $a \wedge b$ denotes the minimum of $a$ and $b$. Let $m_i$ be the total number of observed recurrent events of subject $i$, i.e. $m_i=N(C_i)=N^*(C_i)$.  Define $\bX_i=(1, \tilde\bX_i\trans)\trans$, where $\tilde\bX_i$ is a $(p-1)\times 1$ vector capturing  baseline covariates.  
The observed data include $\{N_i(t), C_i, \bX_i\}_{i=1}^n$. Hereafter the same notation without subscript $_i$ represent the corresponding population analogues.

As described in Section 1, we assume a general multiplicative intensity model, model \eqref{model1}, for the subject-specific occurrence of recurrent events.  That is,  given $\gamma_i$ which summarizes the risk of recurrent events for subject $i$,  $N_i^*(t)$ is a non-stationary Poisson process with the intensity function,
$\lambda(t|\gamma_i)=\gamma_i\cdot \lambda_0(t)$.
We do not specify any parametric form for $\lambda_0(t)$;  nonetheless, we impose  a constraint,
\begin{equation}\label{constraint1}
\int_0^{\nu^*} \lambda_0(t)dt=1,
\end{equation}
 for the purpose of model identifiability, where $\nu^*$ is a predetermined constant which can be chosen as the upper bound of $C_i$'s support. It is clear that, without constraint \eqref{constraint1}, model \eqref{model1} is not identifiable from an alternative specification of the intensity function, $(\gamma_i/c)\cdot \{c\lambda_0(t)\}$, where $c$ is a positive constant. Thus, constraint \eqref{constraint1} is an integral part of the assumed multiplicative intensity model \eqref{model1}.
 Of note, model \eqref{model1} implies
 $$
 E[N_i^*(t)|\gamma_i]=\gamma_i \cdot \mu_0(t),
 $$
 where $\mu_0(t)=\int_0^t \lambda_0(s)ds$, and $\mu_0(\nu^*)=1$. This suggests an alternative interpretation of $\gamma_i$, which is the subject-specific scale shift in the mean function.

The core component of the proposed modeling is to use quantile regression to explore the heterogeneity in subject-specific risk of recurrent events, quantified by $\gamma_i$. Let $Q_{\gamma_i} ( \tau | \bX_i )$ denote the $\tau$th conditional quantile of $\gamma_i$ given $\bX_i$, i.e. $Q_{\gamma_i} ( \tau | \bX_i ) = \inf \{u \geq 0: \Pr( \gamma_i \leq u | \bX_i ) \geq \tau \}$.
We assume that
\begin{eqnarray}  \label{model2}
Q_{\gamma_i} ( \tau | \bX_i ) = \exp\{ \bX_i^{\T} \bbeta_0(\tau) \}\doteq\exp\{a_0(\tau)+\tilde\bX_i\trans\bb_0(\tau)\},
\end{eqnarray}
where $\bbeta_0(\cdot)\doteq(a_0(\tau), \bb_0(\tau)\trans)\trans$ is a $p \times 1$ vector of unknown regression coefficient functions. The non-intercept coefficients in $\bb_0(\tau)$ represent the effects of the corresponding covariates on the $\tau$-th quantile of $\gamma_i$.

Under traditional multiplicative intensity modeling \citep[e.g.]{Nielsen1992, Oakes1992, Wang2001}, $\gamma_i$ is essentially specified by a log-linear model, $\log\gamma_i=\tilde\bX_i\trans \bb_0+\log\xi_i$, where the exponentiated error term, $\xi_i$, corresponds to the so-called frailty. This set-up is a special case of model \eqref{model2} with $\bb_0(\tau)$ being the constant $\bb_0$ and $a_0(\tau)$ equals the $\tau$-th quantile of $\log \xi_i$. If $a_0(\tau)$ is a constant $a_0$, then $\log\gamma_i$ degenerates to $a_0+\tilde\bX_i\trans \bb_0$ and the proposed model reduces to the proportional intensity model \citep{Andersen1982}.  Such connections with the existing models suggest that the non-constancy of $\bbeta_0(\tau)$, as permitted by model \eqref{model2}, accounts for sources of  heterogeneity in $\gamma_i$ not captured by the location-shift effects of the observed covariates \citep{Portnoy2003}. The general multiplicative intensity formulation in \eqref{model1}--\eqref{constraint1}, coupled with the quantile regression modeling of $\gamma_i$ in \eqref{model2}, represents a broader view to accommodate the unobservable subject-specific variability or frailty of recurrent events.

Based on the proposed models \eqref{model1}-\eqref{model2}, we adopt the following censoring assumptions:\\
(i): $C_i$ is independent of $N_i^*(\cdot)$ given $\gamma_i$;\\
 (ii): $C_i$ is independent of $\gamma_i$ given $\tilde\bX_i$. \\
 These assumptions  allows $C_i$ to depend $\tilde\bX_i$. 
 In the special case of $\log\gamma_i=\tilde\bX_i\trans\bb_0+\log\xi_i$,  the assumption (ii) is equivalent to assuming $C_i$ is independent of the frailty $\xi_i$ given $\tilde\bX_i$.

\section{Estimation and Inference}

\subsection{Estimating equation}

It is easy to see that model (\ref{model2}) is equivalent to
$
Q_{ \log(\gamma_i) } (\tau | \bX_i) = \bX_i^{\T} \bbeta_0(\tau),
$
where $Q_{ \log(\gamma_i) } (\tau | \bX_i)$ denotes the $\tau$th conditional quantile of $\log(\gamma_i)$ given $\bX_i$.  If $\gamma_i$'s were observed, we can easily estimate $\bbeta_0(\tau)$ through the score equation of the classic quantile loss function \citep{Koenker1978}:
\begin{eqnarray}  \label{eq1}
 \sum_{i=1}^n  \bX_i \cdot \psi_{\tau} \{ \log( \gamma_i ) - \bX_i^{\T} \bb \}  = 0,
\end{eqnarray}
where $\psi_{\tau} (v)= \tau - I(v < 0)$, $I(\cdot)$ denotes the indicator function, and $\bb \in \mathbb{R}^p$ is a $p$-dimensional unknown coefficients.

A key challenge for estimating $\bbeta_0(\tau)$ is that $\gamma_i$'s are not observable. 
A naive approach to address this difficulty is to replace the $\gamma_i$ in \eqref{eq1} by its observable proxy. Since model \eqref{model1} implies $E[N_i^*(t)|\gamma_i]=\gamma_i\mu_0(t)$, an intuitive proxy of $\gamma_i$ is given by
$\hat\gamma_i=m_i/\hat\mu(C_i)$, where $\hat\mu(\cdot)$ is an estimator of $\mu_0(\cdot)$. An example of $\hat\mu(\cdot)$ is discussed at the end of this subsection. However, this naive approach, as evidenced by our simulation studies, can produce considerably biased estimation by  ignoring the non-negligible deviation of $\hat\gamma_i$ from $\gamma_i$ on the subject-level, despite that $n^{-1}\sum_{i=1}^n \hat\gamma_i$ consistently estimate $E(\gamma_i)$.

Our strategy to deal with unobservable $\gamma_i$'s
is to apply the principle of conditional score \citep{Stefanski1987} to transform the score equation \eqref{eq1} that involves unobservable $\gamma_i$'s to a valid estimating equation that only uses observable quantities. Specifically, we utilize the fact that
\begin{eqnarray}
0 &=& E \left[ \bX \cdot \psi_{\tau} \{ \log(\gamma) - \bX^{\T} \bbeta_0(\tau) \}  \right]  \nonumber \\
  &=& E \left[ E \left\{ \bX \cdot \psi_{\tau} \{ \log(\gamma) - \bX^{\T} \bbeta_0(\tau) \} | m, C, \bX \right\}  \right] \nonumber \\
  &=& E \left\{  \int_{r} \bX \cdot \psi_{\tau} \{ \log(r) - \bX^{\T} \bbeta_0(\tau) \} \cdot f \{ r|m, C, \bX; \bbeta_0(\cdot), \mu_0(\cdot) \} d r   \right\},\label{fact1}    \qquad
\end{eqnarray}
where $f\{ \gamma|m, C, \bX; \bbeta_0(\cdot), \mu_0(\cdot) \}$ denotes the conditional density of $\gamma$ given $m$, $C$ and $\bX$, which depends on the true coefficient function $\bbeta_0(\cdot)$ and the true baseline cumulative intensity function $\mu_0(\cdot)$. The above equation reflects a critical idea that we choose  $(m, C)$ as the surrogate data to  recover the information on $\gamma$.  As elaborated later, such a choice brings analytical convenience as well as computational feasibility. 

Motivated by equation \eqref{fact1}, we consider constructing an estimating equation based on
$$
\bS_n( \bbeta, \mu, \tau ) \doteq \frac{1}{n} \sum_{i=1}^n \int_{r} \bX_i \cdot \psi_{\tau} \{ \log(r) - \bX_i^{\T} \bbeta( \tau ) \}  f\{r | m_i, C_i,  \bX_i; \bbeta(\cdot), \mu(\cdot) \} d r.
$$
It is clearly seen from \eqref{fact1} that $E[ \bS_n( \bbeta_0, \mu_0, \tau ) ]= 0$.

To utilize $\bS_n( \bbeta, \mu, \tau )$ to estimate $\bbeta_0(\tau)$, a  crucial step is to  derive the analytic form of $f\{\gamma |  m, C,  \bX; \bbeta_0(\cdot), \mu_0(\cdot) \}$.  To this end, we note that
\begin{eqnarray}\label{density0}
  f\{\gamma | m, C, \bX; \bbeta_0(\cdot), \mu_0(\cdot)\} &=& \frac{ \rho\{m| \gamma, C, \bX; \mu_0(\cdot)\}g\{ \gamma |C, \bX; \bbeta_0(\cdot) \} }
         { \int_{r}  \rho\{m| r, C, \bX; \mu_0(\cdot)\} g\{ r |C, \bX; \bbeta_0(\cdot) \} d r}    \nonumber   \\
  &=&  \frac{ \rho\{m| \gamma, C; \mu_0(\cdot)\} g\{ \gamma |\bX; \bbeta_0(\cdot) \} }{ \int_{r}  \rho\{m| r, C; \mu_0(\cdot)\} g\{ r |\bX; \bbeta_0(\cdot) \} d r}.
\end{eqnarray}
where ${\rho\{m| \gamma, C, \bX; \mu_0(\cdot)\}}$ denotes the conditional probability mass function of $m$ given $(\gamma, C, \bX)$ and  $g\{ \gamma | C, \bX; \bbeta_0(\cdot) \}$ denotes the conditional density of $\gamma$ given $(C, \bX)$.  The censoring assumption (ii) implies that $g\{ \gamma |C, \bX; \bbeta_0(\cdot) \} $ is free of $C$ and so we can simplify the notation $g\{ \gamma |C, \bX; \bbeta_0(\cdot) \} $ to $g\{ \gamma | \bX; \bbeta_0(\cdot) \} $. We can also omit $\bX$ from ${\rho\{m| \gamma, C, \bX; \mu_0(\cdot)\}}$ because $m=N^*(C)$ and thus its distribution is fully determined when  $\gamma$ and $C$ are given. These justify  the second equality in \eqref{density0}.

%

First, we examine ${\rho\{m| \gamma, C; \mu_0(\cdot)\}}$ using the fact that under model (\ref{model1}), $N^*(t)$, given $\gamma$,  is a nonhomogeneous Poisson process with mean function $\gamma \mu_0(t)$ \citep{Lin2000}. 
This implies that $\{ \mu_0(T^{(1)}), \mu_0(T^{(2)}), \ldots \}$ can be viewed as random variates generated from a homogeneous Poisson process with mean function $\gamma t$. Using standard probabilistic arguments, we show  in
Appendix A that, for both $m=0$ and $m>0$,
\begin{equation}\label{density1}
{\rho\{m| \gamma, C; \mu_0(\cdot)\}} =  \frac{\{\gamma  \mu_0(C)\}^m} { m!}  { \exp\{ -\gamma \mu_0(C)\} }.
\end{equation}

Next, we assess $g\{ \gamma | \bX; \bbeta_0(\cdot) \}$ based on the relationship between the conditional density function and the conditional quantile function of $\gamma$, following the idea of \cite{Wei2009}. That is, given the conditional quantile function specified under  model \eqref{model2},  we can write
\begin{equation}
\label{density2}
g\{ \gamma | \bX; \bbeta_0(\cdot) \} = \lim_{ \delta \rightarrow 0 } \frac{ \delta }{ \exp\{ \bX^{\T} \bbeta_0( \tau_{\gamma} + \delta ) \} - \exp\{ \bX^{\T} \bbeta_0( \tau_{\gamma} ) \} },
\end{equation}
where $\tau_{\gamma} = \{ \tau \in (0, 1): \exp\{ \bX^{\T} \bbeta_0(\tau) \} = \gamma \}$.
Using the results in \eqref{density0}, \eqref{density1}, and \eqref{density2}, we can express  the $f\{\gamma | m_i, C_i,  \bX_i; \bbeta_0(\cdot){, \mu_0(\cdot)} \}$  explicitly in terms of $\gamma$, {$m_i$,} $C_i$, $\bX_i$, $\mu_0(\cdot)$, and $\bbeta_0(\cdot)$.

To construct an estimating equation based on $\bS_n( \bbeta, {\mu,} \tau )$, there remains a major obstacle, which is the unknown infinitely-dimensional {$\mu(\cdot)$}. To address this difficulty, one may follow the conditional likelihood arguments in \cite{Wang2001} to obtain a nonparametric estimator of $\mu_0(t)$, which has a simple product-limit representation. Alternatively, in the following we derive an asymptotic equivalent estimator of $\mu_0(\cdot)$, which takes the Nelson-Aalen form.  
Define the functions $S_C(t|\gamma_i)\doteq \Pr(C\geq t|{\gamma_i})$ and $H_0(t)\doteq\log\{\mu_0(t)/\mu_0(\nu^*)\}$. Given the constraint \eqref{constraint1}, $\mu_0(\nu^*)=1$, it is easy to see that $H_0(\nu^*)=0$ and $\mu_0(t)=\exp\{H_0(t)\}$.  Under the censoring assumption (i) that $ C_i$ is independent of $N_i^*(\cdot)$ given $\gamma_i$, the multiplicative intensity structure imposed by model \eqref{model1} implies that
$$
E\{dN_i(t)|\gamma_i\}=S_C(t|\gamma_i) E\{dN_i^*(t)|\gamma_i{\}}=S_C(t|\gamma_i) \gamma_i\lambda_0(t)dt,
$$
and
$$
E\{I(C_i\geq t)N_i(t)dH_0(t){|\gamma_i}\}=S_C(t|{\gamma_i}) \gamma_i\mu_0(t)\frac{\lambda_0(t)}{\mu_0(t)} dt=S_C(t|{\gamma_i}) \gamma_i\lambda_0(t)dt.
$$
It then follows that $E\{dM_i(t)\}=0$, where $dM_i(t)\doteq dN_i(t)-I(C_i\geq t)N_i(t)dH_0(t)$.
Solving $\sum_{i=1}^n dM_i(t)=0$ yields an estimator of $\mu_0(t)$, which is given by $\hat\mu(t)=\exp\{\hat H(t)\}$ with
$$
\hat H(t)=-\int_t^{\nu^*}\frac{\sum_{i=1}^n dN_i(s)}{\sum_{i=1}^n I(C_i\geq s)N_i(s)}.
$$
Plug $\hat\mu(t)$ into the explicit expression of $f\{\gamma|m_i, C_i, \bX_i; \bbeta(\cdot){, \mu(\cdot)}\}$ and denote the resulting  $f\{\gamma|m_i, C_i, \bX_i; \bbeta(\cdot){, \mu(\cdot)}\}$ and $\bS_n( \bbeta, {\mu,} \tau )$ by ${f}\{\gamma|m_i, C_i, \bX_i; \bbeta(\cdot){, \hat{\mu}(\cdot)}\}$  and ${\bS_n}( \bbeta, { \hat{\mu},} \tau )$ respectively. Then the proposed estimating equation takes the form
\begin{equation}\label{eq2}
n^{1/2} {\bS_n}( \bbeta, { \hat{\mu},} \tau )=0.
\end{equation}
We shall derive the proposed estimator of $\bbeta_0({\tau})$, denoted by $\hat\bbeta(\tau)$, from this estimating equation.


%

\subsection{Estimation algorithm}  \label{ssea}
Solving estimating equation  \eqref{eq2} is not straightforward because $g\{\gamma|\bX;\bbeta(\tau){\}}$ as  in \eqref{density2}  is expressed as a limit and ${\bS_n}(\bbeta, {\hat{\mu},} \tau) $ involves integrals with respect to $\gamma$. In the following we present a detailed algorithm for {obtaining}  $\hat\bbeta(\cdot)$.

First, we assess $g\{\gamma|\bX; \bbeta(\cdot)\}$, which is expressed as a limit as in \eqref{density2}, by adapting the strategy proposed by \cite{Wei2009} for quantile regression with covariate measurement errors. Specifically,  we approximate $\bbeta(\tau)$ by using splines with $K(n)$ knots, $\mathcal{S}_{K(n)} = \{ 0=\tau_0 < \tau_1 < \tau_2 < \ldots < \tau_{K(n)} < 1 \}$.  For a smooth function on $(0, 1)$, the difference between its spline approximation and itself is negligible when the number of knots $K(n)\rightarrow \infty$ as $n\rightarrow\infty$.  For brevity, we shall use notation $K$ instead of $K(n)$ hereafter. We consider  the following piecewise-linear spline approximation,
$$
\sum_{k=1}^K \left[ \bbeta(\tau_{k-1}) + \frac{ \tau - \tau_{k-1} }{ \tau_{k} - \tau_{k-1} } \left\{ \bbeta( \tau_{k} ) - \bbeta( \tau_{k-1} )   \right\}\right]I(\tau_{k-1} < \tau \leq \tau_{k}),
$$
for all $k=1, \ldots, K-1$.
Then, given \eqref{density2}, we may approximate $g\{ \gamma | \bX; \bbeta(\cdot) \} $ by
\begin{eqnarray*}
\tilde g( \gamma | \bX; \btheta ) &=& \sum_{k=1}^K \frac{ \tau_{k} - \tau_{k-1} }{ \exp\{ \bX^{\T}\bbeta ( \tau_{k} ) \} - \exp \{ \bX^{\T}\bbeta(\tau_{k-1}) \} }   \\
                                  & & ~~~~~~~~ \cdot I \{ \exp \{ \bX^{\T}\bbeta(\tau_{k-1}) \} < \gamma \leq \exp\{ \bX^{\T}\bbeta ( \tau_{k} ) \} \},
\end{eqnarray*}
where $\btheta = (\bbeta(\tau_1){\trans}, \bbeta(\tau_2){\trans}, \ldots, \bbeta(\tau_K){\trans} )\trans$ is a ${ (K \cdot p)}$-dimensional parameter, and $\exp\{\bX\trans\bbeta(0)\}$ is fixed as $0$.

Let $\tilde f( \gamma | m_i, $ $ C_i, \bX_i; \btheta) $ denote  ${f}\{ \gamma | m_i, $ $ C_i, \bX_i; \bbeta(\cdot){, \hat{\mu}(\cdot)} \} $ with $\tilde g(\gamma | \bX; {\btheta})$  in place of $g\{ \gamma | \bX; \bbeta(\cdot) \}$. 
Using $\tilde f(\gamma | m_i, $ $ C_i, \bX_i; {\btheta}) $ in place of ${f}\{ \gamma | m_i, $ $ C_i, \bX_i; \bbeta(\cdot){, \hat{\mu}(\cdot)} \} $, we transform equation \eqref{eq2} into an estimating equation,  which can be written as  
\begin{eqnarray} \label{eq3}
\bS_n( \btheta ) \doteq  \frac{1}{n} \sum_{i=1}^n \int_{\gamma} \Psi \{ \log(\gamma) - \bX_i^{\T} \btheta \}  \otimes \bX_i \cdot  \tilde f(\gamma | m_i, C_i, \bX_i; \btheta ) d \gamma = 0,
\end{eqnarray}
where  $\Psi \{ \log(\gamma) - \bX_i^{\T} \btheta \} =(\psi_{\tau_1} \{ \log(\gamma) - \bX_i^{\T} \bbeta(\tau_1) \}, \ldots, \psi_{\tau_K} \{ \log(\gamma) - \bX_i^{\T} \bbeta(\tau_K) \} )^{\T}$, and $\otimes$ denotes Kronecker product. The new estimating equation \eqref{eq3} only involves ${(K \times p)}$ unknown parameters in $\btheta$.

Based on equation \eqref{eq3}, we develop the following algorithm for estimating $\bbeta_0(\cdot)$:

 {\it Step 1.} Set the initial value $\btheta^{[0]}=(\hat{\bbeta}^{[0]} (\tau_1){\trans}, \ldots, \hat{\bbeta}^{[0]} ( \tau_K ){\trans} )\trans$ as the naive estimates obtained from solving a standard quantile regression problem in equation \eqref{eq1} with $\hat\gamma_i$ replacing $\gamma_i$. Set $r=1$.

 {\it Step 2.} Based on $\btheta^{[r-1]}$, evaluate 
        $$
        f^{[r]}(\gamma| m_i, C_i, \bX_i; \btheta^{[r-1]}) = \frac{{\rho\{m_i|\gamma, C_i; \hat{\mu}(\cdot)\}} \tilde g(\gamma|\bX_i; \btheta^{[r-1]}) }{ \int_{\gamma} {\rho\{m_i|\gamma, C_i; \hat{\mu}(\cdot)\}} \tilde g(\gamma|\bX_i;\btheta^{[r-1]}) d \gamma },
        $$
        where
        $$
        {\rho\{m_i|\gamma, C_i; \hat{\mu}(\cdot)\}}=\frac{\{\gamma\hat\mu(C_i)\}^{m_i}}{m_i!}\exp\{-\gamma\hat\mu(C_i)\}.
        $$
%
 {\it Step 3.} Update $\btheta^{[r]}=(\hat{\bbeta}^{[r]}(\tau_1){\trans}, \ldots, \hat{\bbeta}^{[r]}(\tau_K){\trans})\trans$ by the solution to \eqref{eq3} with   $\tilde f(\gamma| m_i, C_i, \bX_i)$ evaluated at $f^{[r]}(\gamma| m_i, C_i, \bX_i)$. Increase $r$ by 1.

 {\it Step 4.} Repeat Steps 2 and 3 until the algorithm converges.

  {\it Step 5.} The proposed estimator is given by
  $$
  \hat\bbeta(\tau) = \sum_{k=1}^K \left[\hat\bbeta(\tau_{k-1}) + \frac{ \tau - \tau_{k-1} }{ \tau_{k} - \tau_{k-1} } \left\{\hat\bbeta( \tau_{k} ) - \hat\bbeta( \tau_{k-1} )   \right\}\right] I(\tau_{k-1} < \tau \leq \tau_{k}).
  $$

To implement the presented algorithm, we adopt numerical integration to assess the integrals with respect to $\gamma$.  In Step 3, finding the solution to \eqref{eq3} can be transformed to a weighted quantile regression problem.
More specifically, let $\tilde{\bgamma}_i^{[r]} = (\tilde{\gamma}_{i, 1}^{[r]}, \tilde{\gamma}_{i, 2}^{[r]}, \ldots, \tilde{\gamma}_{i, J}^{[r]} )$ be a fine grid of possible $\gamma_i$ values in the $r$th step. The estimating equations (\ref{eq3}) can be approximated by
\begin{eqnarray}  \label{eq4}
\sum_{i=1}^n \sum_{j=1}^{J-1} \bX_i \cdot \psi_{\tau_k} \{ \log( \tilde{\gamma}_{i,j}^{[r]} )-\bX_i^{\T} \bbeta(\tau_k) \}  f^{[r]}(\tilde{\gamma}_{i, j}^{[r]} | m_i, C_i, \bX_i)
        ( \tilde{\gamma}_{i, j+1}^{[r]} - \tilde{\gamma}_{i, j}^{[r]} ) = 0,
\end{eqnarray}
for $k=1, \ldots, K$. This can be viewed as a weighted quantile regression problem with responses, $\log( \tilde{\gamma}_{i,j}^{[r]} )$, and  covariates, $\bX_i$, along with weights $f^{[r]}(\tilde{\gamma}_{i, j}^{[r]}\ | m_i, C_i, \bX_i) ( \tilde{\gamma}_{i, j+1}^{[r]} - \tilde{\gamma}_{i, j}^{[r]} )$. Then estimating equations (\ref{eq4}) can be solved by standard statistical software, such as the $rq()$ function in R package $quantreg$.

%
%

The presented estimation algorithm involves the choice of the $\tau$-grid $\mathcal{S}_{K}$  and the $\gamma$-grid $\tilde{\bgamma}_{i}^{[r]}$. As shown by our asymptotic studies,  we require the grid size of  $\mathcal{S}_{K}$,  defined as $\| \mathcal{S}_K \|\doteq \max\{\tau_{k+1}-\tau_k, ~k=1, \ldots, K-1\}$, is of asymptotic order $o(n^{-1/2})$. 
We also suggest
choosing $J=K$ and setting $\tilde{\bgamma}_i^{[r]}$  as $\{\exp(\bX_i^{\T} \hat{\bbeta}^{[r-1]}(\tau_1) ), \ldots, \exp( \bX_i^{\T} 
 \hat{\bbeta}^{[r-1]}(\tau_K) ) \}$ for computational simplicity.

%

\subsection{Large sample properties}

We study the asymptotic properties of  the proposed estimator $\hat{\bbeta}(\cdot)$. For a vector $\bu$, denote $\| \bu \|$ as its Euclidean form. 
Define ${ \bs( \bbeta, \mu, \tau ) = E[ \bS_n( \bbeta, \mu, \tau ) ]}= E [ \int_{\gamma} \bX \cdot \psi_{\tau} \{ \log(\gamma) -
 \bX^{\T} \bbeta(\tau) \} {f \{ \gamma | m, C, \bX; \bbeta(\cdot), \mu(\cdot) \} d \gamma ]}$, then ${ \bs(\bbeta, \mu_0, \tau)} = \tau \bX - E[ \int_{ \gamma } \bX \cdot  
  I \{ \log(\gamma) - \bX^{\T} \bbeta(\tau) < 0 \} \cdot {f \{ \gamma | m, C, \bX; \bbeta(\cdot), \mu_0(\cdot) \} d \gamma ] \doteq \tau \bX - \bupsilon(\bbeta)}$. Define
$h_{\bX}( \tau ) = \{ (\exp\{ \bX^{\T}\bbeta_0(\tau) \})^\prime \}^{-1} = \exp\{ -\bX\trans\bbeta_0(\tau)\} \{\bX\trans \bbeta_0^\prime(\tau)\}^{-1}.$
Note that $h_{\bX}( \tau )$ stands for the density of $\gamma$ given $\bX$ at its $\tau$th conditional quantile, which is $\exp\{ \bX^{\T}\bbeta_0(\tau)\}$ under model \eqref{model2}.  Define
{ $\mathcal{U} = \{ \mu: [0, \nu^*] \rightarrow [0, 1], \mu(\cdot) $ is a nonnegative continuous increasing function with $\mu(0)=0$ and $\mu(\nu^*)=1$\},  }
 $ \mathcal{G} = \{ \bbeta:[\tau_1, \tau_K] \rightarrow \mathbb{R}^p, \bbeta(\cdot)$ is a piecewise linear function with knots in $ \mathcal{S}_K \}$. Let $\mathcal{D}$ be a function space that contains all continuous functions mapping [0, 1] to $\mathbb{R}^p$. Let $\mathcal{X}$ denote the set containing all possible values of $\bX$.

The following are the regularity conditions:
\\
C1. (a) $\gamma$ has a bounded support;
(b) $\mathcal{X}$ is compact; (c) { For any $\mu \in \mathcal{U}$, $\rho\{ \gamma|m, C; \mu(\cdot) \}$} is bounded away from zero and infinity for all $(\gamma, m, C)$;
(d) $g\{ \gamma | \bX; \bbeta(\cdot) \}$ is continuous, bounded away from zero and infinity for all $(\gamma, \bX)$, and $\bbeta \in \mathcal{G}$.\\[2mm]
C2. $\bbeta_0(\tau)$ is Lipschitz continuous in $\tau \in (0, 1)$.
\\[2mm]
C3. (a) $\bupsilon( \bbeta )$, as a functional of $\bbeta(\cdot)$ defined on $\mathcal{D}$, is Fr$\acute{\mbox{e}}$chet differentiable at $\bbeta_0(\cdot)$ with continuously invertible derivative $\dot{\bupsilon}_{\bbeta_0}$; (b) $\| \dot{\bupsilon}_{\bbeta_0}(\bh) \| > 0$ for any $\bh \in \mathcal{F}$ such that $\sup_{\tau \in (0, 1)} \| \bh(\tau) \| \neq 0$, where $\mathcal{F} = \{ c(\bG_1 - \bG_2): c \in \mathbb{R}, \bG_j \in \mathcal{D}, j=1,2 \}$.
\\[2mm]
C4. (a) For any $\bX\in\mathcal{X}$, $h_{\bX}(\tau)$ is finite, and $\lim_{\tau \rightarrow 0} h_{\bX}(\tau) = \lim_{\tau \rightarrow 1} h_{\bX}(\tau) = 0$;
  (b) The first derivative $h_{\bX}^\prime(\tau)$ is bounded for every $\bX\in\mathcal{X}$.

{ Condition C1 imposes realistic boundedness assumptions for $\gamma$, $\bX$, and $g\{ \gamma | \bX; \bbeta(\cdot) \}$.
Note, the boundedness assumption for $\gamma$ implies that the observed number of recurrent events, $m$, is bounded. Bounded $m$, coupled with bounded $\mu(\cdot)$, would imply condition C1(c).}
Condition C2 assumes the smoothness of the true coefficient function $\bbeta_0(\tau)$, which has been commonly adopted in quantile regression literature. 
Condition C3 is a key assumption to ensure the identifiability of $\bbeta_0(\tau)$, which resembles those  used in Z-Estimators \citep{Kosorok2008, Ji2014}.  Condition C4 imposes assumptions on $h_{\bX}(\tau)$, which helps justify the approximation of $g\{\gamma|\bX; \bbeta(\cdot)\}$ by $\tilde g(\gamma|\bX; \btheta)$ and was similarly adopted by  \cite{Wei2009}.
Condition C4 accommodates common distributions, such as the Gaussian, Exponential, and Student $t$ distributions.
Given the relationship between the density function and the quantile function, condition C4(a) implicitly implies that $g\{ \gamma | \bX; \bbeta_0(\cdot) \}$ is continuous, bounded away from zero and infinity for all $\gamma$ and $\bX$ .

We establish the uniform consistency and weak convergence of the proposed estimator $\hat\bbeta(\tau)$ in the following theorems:
\begin{theorem}   \label{theo1}
Under regularity conditions C1--C4, if $K \rightarrow \infty$, $ \lim_{n \rightarrow \infty} \| \mathcal{S}_K \| = 0$, and $K/n^{\alpha} \rightarrow 0$ for some $\alpha > 0$, then
$$
\lim_{n \rightarrow \infty} \sup_{ \tau \in [\tau_1, \tau_K] }  \| \hat{\bbeta}(\tau) - {\bbeta_0(\tau)} \| \stackrel{p}{\longrightarrow} 0.
$$
\end{theorem}

\begin{theorem}    \label{theo2}
Under regularity conditions C1--C4, if $K \rightarrow \infty$, $\lim_{n \rightarrow \infty} n^{1/2} \| \mathcal{S}_K \| = 0$, and $K/n^{\alpha} \rightarrow 0$ for some $\alpha > 0$, then $n^{1/2} \{ \hat{\bbeta}(\tau) - \bbeta_0(\tau) \}$ converges weakly to $ \dot{\bupsilon}_{\bbeta_0}^{-1} \{ \bG(\tau) \}$, where {$\bG(\tau)$ is a tight Gaussian process with mean {\bf 0} and covariance function $E [ \boldsymbol{\eta}_i(s) \boldsymbol{\eta}_i(t)^{\T} ]$ for $s, t \in [\tau_1, \tau_K]$, $\boldsymbol{\eta}_i(t) = \boldsymbol{\eta}_{1i}(t) + \boldsymbol{\eta}_{2i}(t)$ with $\boldsymbol{\eta}_{1i}(t)$ and $\boldsymbol{\eta}_{2i}(t)$ defined in the Appendix B.}
\end{theorem}

Note that Theorems 1-2 are focused on the asymptotic properties of $\hat{\bbeta}(\tau)$ with $\tau \in [\tau_1, \tau_K]$, a closed subset of $(0, 1)$. This is necessary because model (\ref{model2}) implies that $\exp\{ \bX^{\T} \bbeta_0(0) \} = 0$ and hence $\|\bbeta_0(0)\}\| = \infty$. 
Nonetheless, $[\tau_1, \tau_K]$ nearly covers the interval $(0, 1)$ given $\Pr( \gamma \leq \exp\{\bX^{\T} \bbeta_0(\tau_1)\} ) = o(1)$ and  $\Pr( \gamma > \exp\{\bX^{\T} \bbeta_0(\tau_K)\} ) = o(1)$. Detailed proofs of Theorems 1-2 are provided in the Appendix B.

\subsection{Bootstrapping-based inference}

To make inference about $\bbeta_0(\tau)$, a bootstrapping procedure may be preferred provided the complexity of the asymptotic covariance matrix derived in the proof of Theorem 2. Specifically, we may resample the observed data with replacement and obtain an estimator of $\bbeta_0(\tau)$ based on the resampled sample, denoted by $\bbeta^*(\tau)$. Repeating this procedure many times can generate a large number of realizations of  $\bbeta^*(\tau)$.   For a fixed $\tau_* \in [\tau_1, \tau_K]$, the variance of $\hat{\bbeta}(\tau_*)$ can be estimated by the empirical variance of $\bbeta^*({\tau_*})$.
The confidence intervals for $\bbeta_0(\tau_*)$ can be constructed using a normal approximation or by referring to the empirical percentiles of $\bbeta^*({\tau_*})$.

In addition, one may be interested in testing whether the non-intercept components and/or the intercept component of $\bbeta_0(\tau) $ are constant over $\tau$ or not. Such hypothesis testing may yield goodness-of-fit tests for existing multiplicative intensity models, including \cite{Andersen1982}'s model and multiplicative intensity models with frailty \citep{Nielsen1992, Oakes1992, Wang2001}. These types of second-stage inference can follow  similar lines of other work on quantile regression \citep[e.g.]{Peng2008, Peng2009}; details are omitted here.

\section{Numerical studies}

\subsection{Monte-Carlo simulations}

In this {sub}section, we conduct simulation studies to evaluate the finite sample performance of the proposed methods.
Specifically, for subject $i$, we  generate recurrent event times $\{ T_i^{(j)},\ \ j=1, 2, \ldots \}$ from a Poisson process with rate $\gamma_i$. In this case, model \eqref{model1} is met with $\mu_0(t)=t$. 

We first consider the situation where $\gamma_i$ satisifies a log-linear model with homogeneous errors:
\begin{eqnarray} \label{eq10}
\log( \gamma_i ) = \bX_i^{\T} { \bb} + 0.5 \epsilon_i,
\end{eqnarray}
where  { $\bb=(b_0, b_1, b_2)\trans$} and $\bX_i=(1, X_{i,1}, X_{i,2})\trans$. We let $X_{i,1}\sim Unif(0, 1)$, and $X_{i,2}\sim Bernoulli(0.5)$. We let $\epsilon_i$ follow the standard normal distribution, $N(0, 1)$, or the Student's $t$-distribution, $t_3$. In this set up, model \eqref{model2} holds with $\beta^{(1)}(\tau)=b_0+0.5 Q_{\epsilon}(\tau)$, { $\beta^{(2)}(\tau)=b_1$, $\beta^{(3)}(\tau)=b_2$}, where the superscript $^{(k)}$ indicates the $k$th component of a vector, and $Q_{\epsilon}(\tau)$ represents the $\tau$th quantile of $\epsilon$. We generate the censoring time $C_i$ from Unif(2/3, 1), independent of $T_i^{(j)}$ and $\tilde\bX_i$. We set {$b_0=\log(3)+1, b_1=b_2=1$}, yielding the average number of observed recurrent events per subject is about 24 or 25.8 corresponding to $N(0,1)$ or $t_3$ error respectively.
Under each configuration, we generate 500 simulated datasets with sample size $n=500$. For each simulated dataset, 100 bootstrapping samples are drawn to calculate the estimated standard error and coverage probability. To implement the proposed method, $\mathcal{S}_K$ is set as an equally spaced grid between $0.02$ and $0.98$ with the size $0.01$. 
We adopt an adjusted version of $\hat\gamma_i$, which is $\max(1, m_i)/\hat\mu_0(C_i)$, to compute the naive estimates for $\bbeta_0(\tau)$.
For the iterative algorithm, the maximum iteration number is set to be 100, and the stop criteria is $ \sum_{k=1}^K \| \bbeta^{[r-1]}(\tau_k) - \bbeta^{[r]}(\tau_k)\|^2 < 0.01$.

\begin{figure}
\centering
\includegraphics[width=12cm, height=12cm]{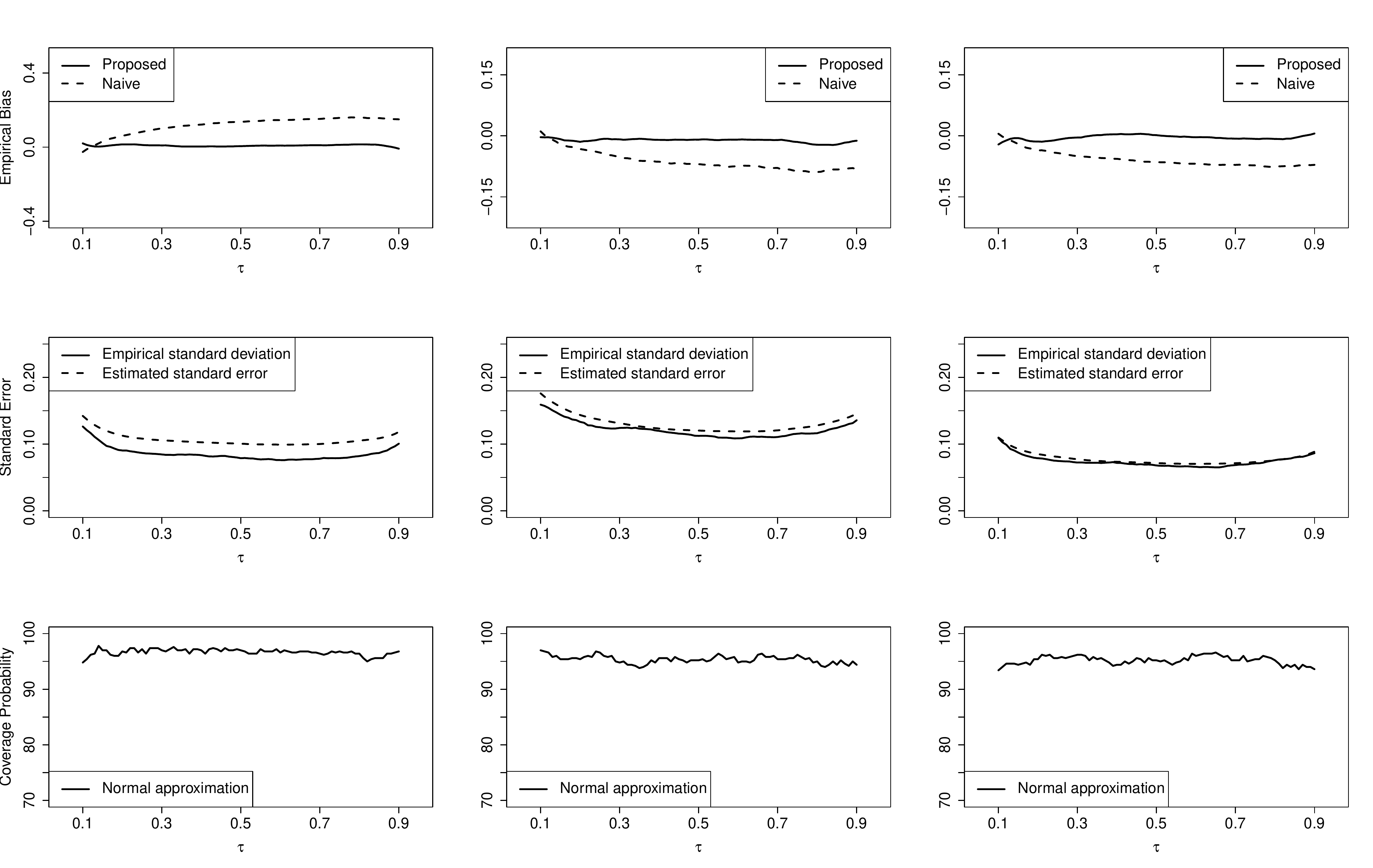}
\caption{ { Simulation results for the homogeneous error case with $\epsilon \sim N(0, 1)$. }
}
\label{fig1}
\end{figure}

\begin{figure}
\centering
\includegraphics[width=12cm, height=12cm]{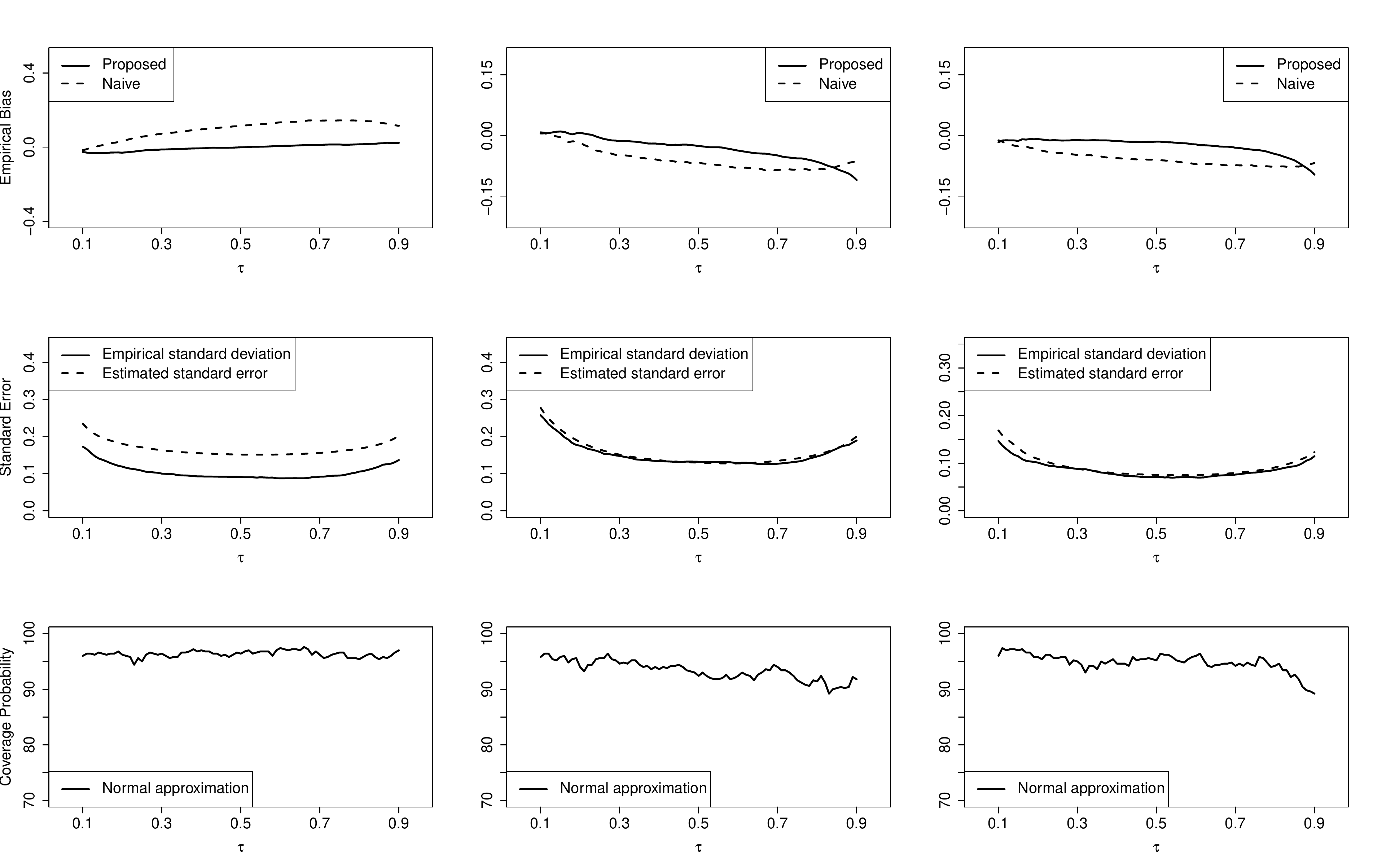}
\caption{ { Simulation results for the homogeneous error case with $\epsilon \sim t_3$. }
}
\label{fig2}
\end{figure}

The simulation results when $\epsilon$ follows $N(0, 1)$ distribution and $t_3$ distribution are summarized in Figures \ref{fig1} and \ref{fig2}, respectively.
The empirical biases of the proposed estimator (solid lines), together with the empirical biases of the naive estimator (dashed lines) are plotted in the first row of each figure.
It is shown that the naive estimator can produce  large biases, especially  for large $\tau$'s.
In contrast, the empirical biases of the proposed estimator are quite small.
The second row of each figure compares  the empirical standard deviations (solid lines) with the estimated standard errors (dashed lines) of the proposed estimator.
We observe that they match with each other very well.
In the third row, we plot the empirical coverage probabilities of the 95\% confidence intervals constructed by normal approximations  that use bootstrapping standard errors. 
The empirical coverage probabilities are reasonably close to the nominal level 95\%.

\begin{figure}
\centering
\includegraphics[width=12cm, height=12cm]{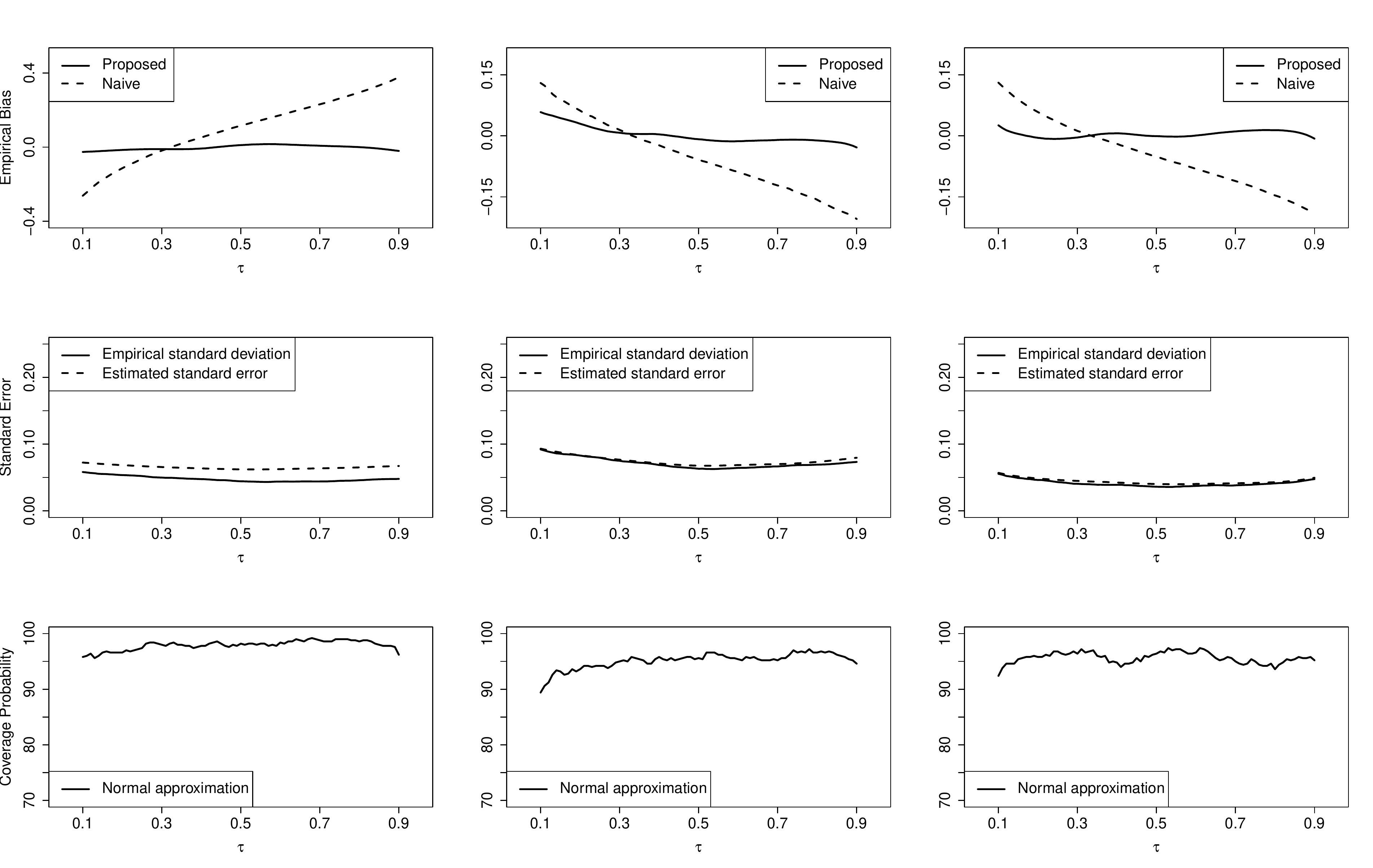}
\caption{ { Simulation results for heteroscedastic case with $\epsilon \sim N(0, 1)$. }
}  \label{fig3}
\end{figure}

We also consider the situation where the non-intercept coefficients in $\bbeta_0(\tau)$ are not constant. To simulate such data, we let $\gamma_i$ follow a  log-linear model with heteroscedastic errors:
\begin{eqnarray}
\log( \gamma_i ) = { \bX_i\trans \bb +  (\bX_i\trans \bd) } \epsilon_i.    \label{req9}
\end{eqnarray}
We generate $\bX_i$ and $C_i$  in the same way as in the first set-up.
We set {$\bb=(b_0, b_1, b_2)\trans$ as before, and set $\bd=(d_0, d_1, d_2)\trans = (0.1, 0.1, 0.1)\trans$.}
We only consider $\epsilon_i$ that follows the standard normal distribution $N(0, 1)$ in this heteroscedastic case.
The average number of observed recurrent events per subject is about 22.8.
Under model (\ref{req9}),  {$\beta^{(i)}(\tau) = b_{i-1} + d_{i-1} Q_{\epsilon}(\tau), i=1,2,3,$} which are changing with $\tau$. 
 In Figure \ref{fig3}, the simulation results  are displayed in the same manner as those in Figures \ref{fig1} and \ref{fig2}.
It is shown that the proposed estimator $\hat{\beta}^{(1)}(\tau)$ has small biases for $\tau$'s ranging from 0.1 to 0.9.
Meanwhile, the biases of $\hat{\beta}^{(2)}(\tau)$ and $\hat{\beta}^{(3)}(\tau)$ are negligible except for that corresponding to extremely small and large $\tau$'s.
The empirical biases of the naive estimator are considerably larger than those of the proposed estimator at most values of $\tau$'s. The observations regarding the variance estimation and the coverage probabilities are similar between Figures \ref{fig1}-\ref{fig2} and Figure \ref{fig3}.  That is, the estimated standard errors agree well with the empirical standard deviations, and the confidence intervals based on normal approximation or percentiles yield quite accurate empirical coverages probabilities.
Overall, our simulation results suggest satisfactory finite-sample performance of the proposed methods.

\subsection{The DURABLE Data Example}
\setcounter{table}{0}

The DURABLE study \citep{Buse2009} is an open-label randomized clinical trial in Type 2 diabetes patients. It was designed to compare the efficacy and safety of two starter insulin regimens, twice-daily lispro mix 75/25 (LM75/25; 75\% lispro protamine suspension, 25\% lispro),  or once-daily glargine (GL), in addition to  oral antihyperglycemic drugs (OADs). This study
enrolled 2,187
insulin-naive patients with type 2 diabetes from 11 countries, aged 30 to 80 years, with HbA1c
$>$ 7.0{$\%$}, and  on at least two oral antihyperglycemic agents.

Hypoglycemia, as an important safety endpoint, was closely monitored during this study.  A large cross-subject variability was noted regarding the occurrences of hypoglycemia during the 24-week study follow-up. For example, the number of hypoglycemia episodes per subject has a wide range,  from 0 to 137, with mean equal to 10.8 and median equal to 5. These descriptive statistics suggest a high degree of heterogeneity in the individual risk of hypoglycemia presented in the DURABLE trial. Exploring the risk factors for hypoglycemia and, moreover, potentially different risk mechanisms between ``frail'' (high risk) versus ``robust'' (low risk) patients are of great clinical interest.
The proposed quantile regression framework for recurrent event data is tailored to address these interests, in particular the latter one, which cannot be addressed by routine recurrent event data analyses.

\begin{table}
\renewcommand{\arraystretch}{1.2}
\centering
\caption{ Summary statistics for the covariates in DURABLE data }
\label{table1}
\begin{tabular}{cccccc}
\hline \hline
   &           & LM75/25  &            & GL  &           \\
 {\it therapy}    &           & 987 (49.3\%)   &            & 1016 (50.7\%)  &      \vspace{0.05in}
      \\
  &           & with TZD   &            & without TZD  &           \\
 {\it tzduse}     &           & 753 (37.6\%)   &            & 1250 (62.4\%)  &           \vspace{0.05in}
 \\
   &           & with SU  &            & without SU   &           \\
 {\it sulfouse}   &           & 1837 (91.7\%)  &            & 166 (8.3\%)    &           \\
\hline
                  &   Median  &   Minimum      &   Maximum  &   Mean         &  Standard Deviation  \\
 {\it basfglu}    &   10.46   &   0.23         &   25.96    &  10.79         &       3.72            \\
 {\it basfins}    &   7.97    &   0.18         &   142.68   &  10.42         &       9.75            \\
 {\it bmibase}    &   31.25   &  15.88         &   62.62    &  31.71         &       6.19            \\
 {\it durdiab}    &   8.50    &   0.03         &   39.48    &  9.73          &       6.17            \\
\hline \\
\end{tabular}
\end{table}

%
%

We apply the proposed method to the DURABLE data. In our analysis,  the recurrent event time $T^{(j)}$ corresponds to the time from study enrollment to the $j$th episode of hypoglycemia.
We consider baseline covariates  including {\it therapy}, which is 1 if the patient had LM75/25 and 0 otherwise,
{\it basfglu}, which represents baseline fasting blood glucose, {\it basfins}, which represents baseline fasting insulin,
{\it bmibase}, which represents baseline body mass index (BMI), {\it durdiab}, which represents duration  of type 2 diabetes,
{\it tzduse}, which is 1 if the patient used thiazolidinedione (TZD) and 0 otherwise, and {\it sulfouse}, which is 1 if the patient used sulfonylurea (SU) and 0 otherwise.
The summary statistics of these covariates are presented in Table 1. 
In our analysis, we standardize the continuous covariates, and exclude subjects with missing covariates or incorrect covariate values (defined by values outside reference range). 
The final sample size is $n=2,003$. We choose $\mathcal{S}_K$ as an equally space grid between 0.02 and 0.98 with step size 0.02. Inferences are carried out based on 200 bootstrapping samples. Other set-ups are the same as those in the simulation studies.

%

\begin{figure}
\centering
\includegraphics[width=12cm, height=12cm]{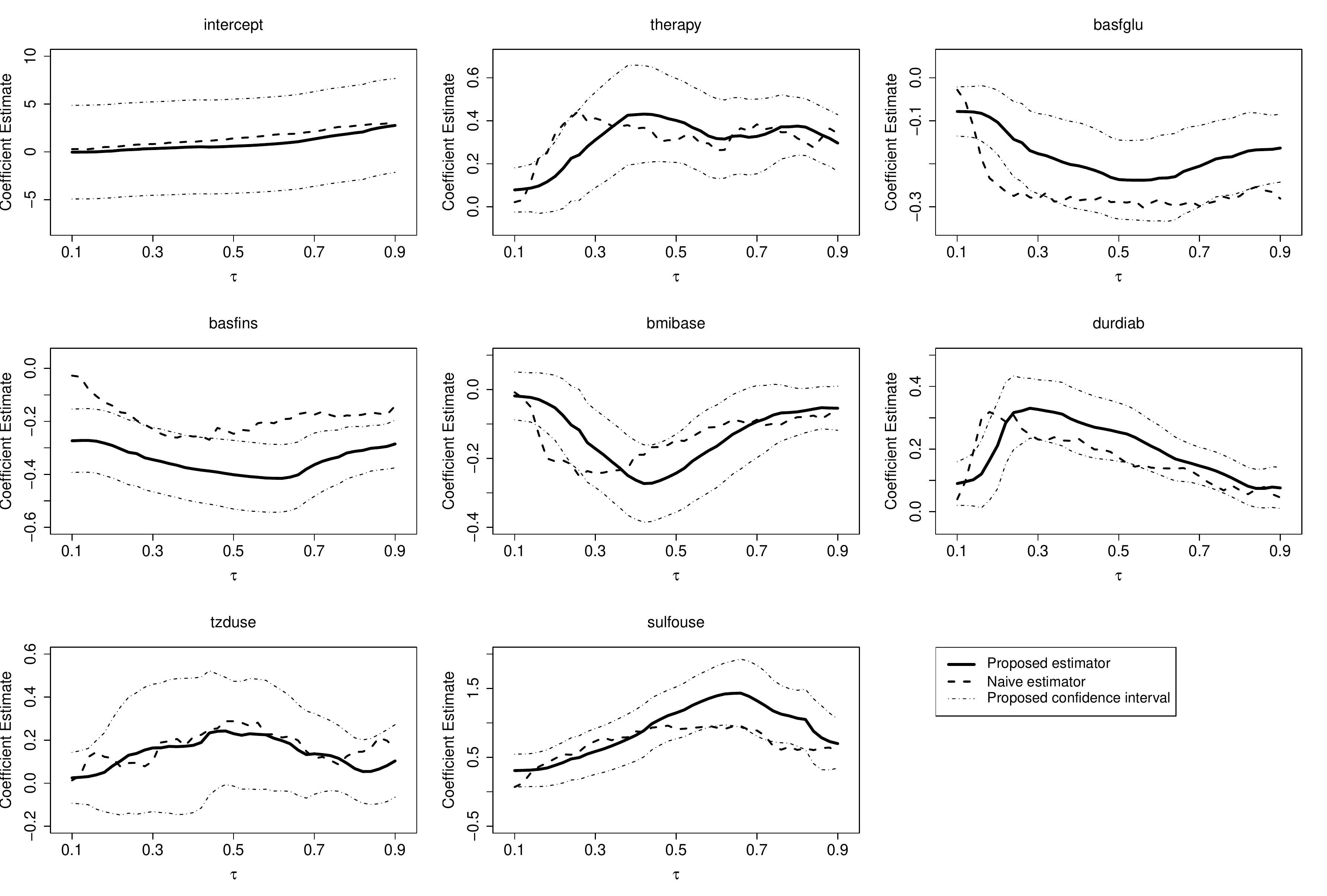}
\caption{The DURABLE data example: the proposed coefficient estimates (red solid lines) and the corresponding pointwise confidence intervals (red dash dotted lines), along with the naive estimates
     (black dashed lines).}  \label{fig4}
\end{figure}

In Figure \ref{fig4}, we plot the estimated coefficients $\hat\bbeta(\tau)$ (red solid lines) and the corresponding 95\% pointwise confidence intervals (red dash dotted lines), along with the naive estimates (black dashed lines) for $\tau \in [0.1, 0.9]$. Under the proposed models \eqref{model1}-\eqref{model2}, positive (or negative) coefficients indicate covariate effects associated with higher (or lower) risk of hypoglycemia, which is measured as subject-specific positive (or negative) scale shift of the intensity or rate function of hypoglycemia recurrence.
We can see from Figure \ref{fig4} that patients receiving LM75/25 demonstrate higher risk of hypoglycemia  than GL patients.
The results in Figure \ref{fig4} also suggest that lower baseline glucose, lower baseline insulin, lower baseline BMI, longer diabetes duration, or using sulfonylurea, are associated with higher risk of hypoglycemia.
Comparing the proposed estimates with the naive estimates, we note that
the naive estimates sometimes show significant departures from the proposed estimates.
For example, the naive estimates for  {\it therapy}'s coefficients are beyond on the upper bound of the confidence intervals when $\tau \in [0.15, 0.25]$. This may be a sign of large estimation bias resulted from using the naive approach, and indicate the need of developing the proposed methods, which are designed to appropriately address the bias issue.

We also conduct second-stage inference to summarize the estimated covariate effects by average covariate effects, defined as
$ \int_{\tau_L}^{\tau_U} \beta_0^{(j)}(\tau)d\tau {/(\tau_U - \tau_L)}$, where $\tau_L=0.1$, $\tau_U=0.9$ and $j=2,\ldots, p$. The inferences on the average effects are conducted by following the lines of \cite{Peng2009}.  We also fit the data with the standard proportional intensity model \citep{Andersen1982}, which is a special case of the proposed models with all coefficients in $\bbeta_0(\tau)$ being constant over $\tau$. In Table 2, 
we present the estimated average effects and the corresponding standard errors and Wald-test $p$-values, along with the coefficient estimates and the corresponding standard errors and $p$-values based on the proportional intensity model. It is seen that the proposed method generates quite consistent findings regarding the impact of covariates on the risk of hypoglycemia. The standard errors based on the proposed method are  larger than those based on the proportional intensity model. This reflects a tradeoff between greater model flexibility and reduced estimation efficiency.    \\

We next employ second-stage inference to test the constancy of each coefficient function in $\bbeta_0(\tau)$. The corresponding null hypothesis is formulated as $H_{0, j}:\  \beta_0^{(j)}(\tau) = \rho_0, ~\tau \in [\tau_L, \tau_U]$,
where $\rho_0$ is an unspecified constant, and $j=2, \ldots, p$.
Following the lines of \cite{Peng2009}, we can use the test statistic, $\mathcal{T} = n^{1/2} \int_{\tau_L}^{\tau_U} \Xi(\tau) \{ \hat{\beta}^{(j)}(\tau) - \hat{ {\eta} }^{(j)} \}  d \tau$,
where $\Xi(\tau)$ is a nonnegative and nonconstant weight function, $\hat{ {\eta} }^{(j)}=\int_{\tau_L}^{\tau_U} \hat{\beta}^{(j)}(\tau)  d \tau /(\tau_U-\tau_L)$.
We choose $\Xi(\tau) = I\{ \tau \leq (\tau_L + \tau_U)/2\}$.
We may reject $H_{0, j}$ if $\mathcal{T} > q_{1-\alpha/2}$ or $\mathcal{T} < q_{\alpha/2}$, where $q_{\alpha/2}$ and $q_{1-\alpha/2}$ are the $\alpha/2$th and $(1-\alpha/2)$th empirical quantiles of
$\mathcal{T}^* = n^{1/2} \int_{\tau_L}^{\tau_U} \Xi(\tau) [ \{ \beta^{*(j)}(\tau) - \hat{\beta}^{(j)}(\tau) \} - \{ {\eta}^{*(j)} - \hat{ {\eta} }^{(j)} \} ] d \tau$.
We apply this test to the coefficient for each covariate.
The results indicate that { {\it durdiab} and {\it sulfouse} have} non-constant effects over $\tau$, while constant effects are adequate for other covariates.
{ The test statistic $\mathcal{T}$ for the coefficient for {\it durdiab} is $0.913$, lying in the rejection region $(-\infty, -0.588)\cup(0.689, +\infty)$.
The test statistic $\mathcal{T}$ for the coefficient for {\it sulfouse} is $-4.885$, lying in the rejection region $(-\infty, -2.331)\cup(2.787, +\infty)$.}
Combined with the observation from Figure \ref{fig4},  this suggests that the elevated risk of hypoglycemia associated with the use of sulfonylurea may be amplified in subjects who are susceptible to frequent hypoglycemia (corresponding to large $\tau$'s). The clinical implication may be that caution is needed for using sulfonylurea in patients who are known or projected to have a high risk of hypoglycemia based on patient history and clinical judgement.  The non-constancy of the coefficients for { {\it durdiab} and} {\it sulfouse} also provide an evidence for the lack of fit of the proportional intensity model to the DURABLE data.

\begin{table}
\caption{Comparison of the Proposed Method with the Proportional Intensity Model}
\label{table2}
\renewcommand{\arraystretch}{1.2}
\begin{tabular}{ccrcccrcc}
\hline \hline
                & & \multicolumn{3}{c}{  { Average Effect}  }      & &  \multicolumn{3}{c}{   Proportional Intensity Model  }     \\
  \cline{3-5}    \cline{7-9}
                & &  Estimate    &  Standard  &  $p$-value   & &  Estimate & Standard & $p$-value \\
                & &              &     Error  &              & &           &   Error  &           \\
\hline
 {\it therapy}  & &   { 0.315 }     &    {   0.072 }     &  $<0.001$    & &   0.274   &    0.015       &  $<0.001$     \\
 {\it basfglu}  & &  { $-0.182$ }   &    {   0.036 }     &  $<0.001$    & & $-0.173$  &    0.008       &  $<0.001$     \\
 {\it basfins}  & &  { $-0.349$ }   &    {   0.052 }     &  $<0.001$    & & $-0.384$  &    0.013       &  $<0.001$     \\
 {\it bmibase}  & &  { $-0.135$ }   &    {   0.042 }     &   0.001      & & $-0.045$  &    0.008       &  $<0.001$     \\
 {\it durdiab}  & &   { 0.201 }     &    {   0.033 }     &  $<0.001$    & &   0.118   &    0.007       &  $<0.001$     \\
 {\it tzduse}   & &   { 0.142 }     &    {   0.092 }     &  0.125       & &   0.113   &    0.016       &  $<0.001$     \\
 {\it sulfouse} & &   { 0.903 }     &    {   0.152 }     &  $<0.001$    & &   0.755   &    0.037       &  $<0.001$     \\
\hline  \\
\end{tabular}
\end{table}

%
%
%
%

\section{Discussion}

In this work, we develop a new quantile regression framework which provides a direct, sensible and flexible approach to addressing the heterogeneity in subject-specific risk of recurrent events. The proposed modeling is built upon a general formulation of the multiplicative intensity model for recurrent events. The recurrent event process is modeled by a subject-specific Poisson process. The event times of the same subject are correlated through sharing the same individual risk measure $\gamma_i$ that may account for either observed covariates or unobservable frailty.  The proposed modeling encompasses several popular recurrent events models, including the proportional intensity model \citep{Andersen1982} and proportional intensity shared frailty model \citep[e.g.]{Nielsen1992, Oakes1992, Wang2001}. Beyond the existing models, the proposed models can accommodate more dynamic relationships between covariates and subject-specific risk of recurrent events that are scientifically meaningful and important. In this work, we propose a new modeling perspective which captures subject-specific risk of recurrent events by the individual scale shift in the intensity or rate function from an unspecified baseline function.  Potentially, the subject-specific risk can be formulated in different ways, for example, by location shifts, which may better some application settings. Such generalizations or variants warrants future research.

\section*{Appendices}

\setcounter{theorem}{0}

\subsection*{Appendix A: The detailed derivation of ${\rho\{m|\gamma, C; \mu_0(\cdot)\}}$}
\renewcommand{\theequation}{A.\arabic{equation}}
\setcounter{equation}{0}

Under model (\ref{model1}), $N^*(t)$, given $\gamma$,  is a nonhomogeneous Poisson process with mean function $\gamma \mu_0(t)$ \citep{Lin2000}. This implies that $\{ \mu_0(T^{(1)}), \mu_0(T^{(2)}), \ldots \}$ can be viewed as random variates generated from a homogeneous Poisson process with mean function $\gamma t$.
Letting  $T^{(0)}=0$, given $\gamma$, we have $\{ \mu_0(T^{(j)}) - \mu_0( T^{(j-1)} ): j=1, 2, \ldots, m \}$ are independent and identically distributed following the exponential distribution, $Exponential(\gamma)$, and  $\mu_0(T^{(m)})$ follows the gamma distribution, $Gamma(\gamma)$, with the density function, $
\frac{ x^{ m-1} }{ (m-1)! } \gamma^{m} \exp(-\gamma x)$.

 When $m=0$,
$
{ \rho\{0| \gamma, C; \mu_0(\cdot)\}  = \Pr\{ m=0|\gamma, C; \mu_0(\cdot) \}}= \Pr( \mu_0(T^{(1)}) > \mu_0(C)| 
 \gamma, C ) = \exp\{-\gamma \mu_0(C)\}.
$
When $m \geq 1$,
\begin{eqnarray*}
{ \rho\{m| \gamma, C; \mu_0(\cdot)\} } &=&  \Pr \left( \mu_0( T^{(m)}) \leq \mu_0(C),\  \mu_0(T^{(m+1)}) > \mu_0(C) | \gamma, C \right)   \\
                   &=&  \Pr \left(0 \leq  \mu_0(C) -  \mu_0( T^{(m)}) < \mu_0(T^{(m+1)}) - \mu_0( T^{(m)}) | \gamma, C \right).
\end{eqnarray*}
Under the censoring assumption (i), we have $C$ is independent of $T^{(j)}$ given $\gamma$. Recall that  $\mu_0(T^{(m+1)}) - \mu_0( T^{(m)})$  and $\mu_0( T^{(m)})$, given $\gamma$, follow $Exponential(\gamma)$ and $Gamma(\gamma)$ distributions respectively. Then,  the  density function of $ \mu_0(C)- \mu_0( T^{(m)} )$ given $(\gamma, C)$ is
$$
\frac{\{\mu_0(C)-x\}^{m-1}}{ (m-1)! } \gamma^m \exp\{ -\gamma \mu_0(C) + \gamma x \},
$$
and the density function of $\mu_0(T^{(m+1)}) - \mu_0( T^{(m)})$ given $(\gamma, C)$ is $\gamma \exp(-\gamma y)$. Hence for  $m>0$, we have
\begin{eqnarray*}
& &  { \rho\{m| \gamma, C; \mu_0(\cdot)\} }  \\ [2mm]
&=&  \int_0^{\mu_0(C)} \frac{\{ \mu_0(C)-x\}^{m-1}}{ (m-1)! } \gamma^m \exp\{ -\gamma  \mu_0(C) + \gamma x \} \left( \int_x^\infty  \gamma \exp(-\gamma y)  dy   \right)  d x   \\ [2mm]
                   &=& \frac{ \{\gamma \mu_0(C)\}^m }{ m! } \exp\{ -\gamma \mu_0(C) \}.
\end{eqnarray*}
This means, for both $m=0$ and $m>0$,
$$
{ \rho\{m| \gamma, C; \mu_0(\cdot)\} } =  \{\gamma  \mu_0(C)\}^m / m! ~{ \exp\{ -\gamma \mu_0(C)\}.}
$$

\subsection*{Appendix B: Proofs of Theorems 1-2}

\begin{lemma} \label{lem0}
Under the regularity condition C1(a), $\hat{\mu}(t)$ is a uniformly consistent estimator of $\mu_0(t)$. Furthermore, $n^{1/2} \{ \hat{\mu}(t) - \mu_0(t) \}$ converges weakly to a zero-mean Gaussian process with covariance function $ E\{ \mu_0(s) \varphi_i(s) \varphi_i(t) \mu_0(t)\}$ at $(s, t)$, where $\varphi_i(t) = - \int_t^{\nu^*} \frac{  d M_i(s) }{ E\{S_C(s|\gamma_i) \gamma_i\} \mu_0(s) } $.
\end{lemma}
{\it Proof of Lemma \ref{lem0}:}
We first note that $m_i=N_i(C_i)$ is bounded by the regularity condition C1(a) and model (\ref{model1}).
We show the uniform consistency and weak convergence of $\hat{H}(t)$ first, and the desired results follow immediately from Taylor expansions and the continuity of $\exp(x)$.

Recall that
$$
\hat{H}(t) = - \int_t^{\nu^*} \frac{ \sum_{i=1}^n d N_i(s) }{ \sum_{i=1}^n I( C_i \geq s ) N_i(s) }.
$$
Consider  $\mathcal{G}_1 = \{ I( C_i \geq s ) N_i(s): s \in [0, \nu^*] \}$ and
$\mathcal{G}_2 = \{ \int_t^{\nu^*} [E\{S_C(s|\gamma_i) \gamma_i\} \mu_0(s)]^{-1}$ $d N_i(s): s \in [0, \nu^*] \}$. It is clear that $\mathcal{G}_1$ is Glivenko-Cantelli because $I( C_i \geq s ) N_i(s)$  only involves indicator functions.  Applying Glivenko--Cantelli theorem to $\mathcal{G}_1$ gives the uniform convergence of $ n^{-1}\sum_{i=1}^n I( C_i \geq s ) N_i(s)$ to $E\{S_C(s|\gamma_i) \gamma_i\} \mu_0(s)$.   According to the regularity condition C1 and model (\ref{model1}),   $E\{S_C(s|\gamma_i) \gamma_i\} \mu_0(s)]^{-1}$ is uniformly bounded and hence the functions in ${\cal G}_2$ is Lipschitz continuous and thus ${\cal G}_2$ is Glivenko-Cantelli. Then, by the Glivenko-Cantelli theorem,
$$
n^{-1}\sum_{i=1}^n -\int_t^{\nu^*}  [E\{S_C(s|\gamma_i) \gamma_i\} \mu_0(s)]^{-1} dN_i(s)
$$
uniformly converges to $- \int_t^{\nu^*} \lambda_0(s) d s/\mu_0(s)$, which equals to $H_0(t)$ because $H_0(\nu^*)=0$.
This, coupled with the uniform convergence of $ n^{-1}\sum_{i=1}^n I( C_i \geq s ) N_i(s)$ to $E\{S_C(s|\gamma_i) \gamma_i\} \mu_0(s)$, implies the uniform convergence of $\hat{H}(t)$ to $H_0(t)$.

To show the weak convergence of $\hat{H}(t)$, write
\begin{eqnarray*}
n^{1/2} \{ \hat{H}(t) - H_0(t) \} &=& - n^{1/2}  \int_t^{\nu^*} \frac{ \sum_{i=1}^n d M_i(s) }{ \sum_{i=1}^n I( C_i \geq s ) N_i(s) }.
\end{eqnarray*}
{
Because each summand in $M_i(t)$ is the monotone function of $t$, $\{ M_i(t): t \in [0, \nu^*] \}$ is a Donsker class \citep{van1996}, and $n^{-1/2} \sum_{i=1}^n M_i(t) = O_p(1)$ uniformly in $t \in [0, \nu^*]$.}
This, coupled with the uniform consistency of
$n^{-1} \sum_{i=1}^n I(C_i \geq s) N_i(s)$, implies that
\begin{eqnarray}
n^{1/2} \{ \hat{H}(t) - H_0(t) \}  &=& - n^{-1/2} \sum_{i=1}^n \int_t^{\nu^*} \frac{  d M_i(s) }{ E\{S_C(s|\gamma_i) \gamma_i\} \mu_0(s) } + \epsilon_0(t)    \nonumber   \\
                                   & \doteq & - n^{-1/2} \sum_{i=1}^n \varphi_i(t) + \epsilon_0(t),               \label{eqh}
\end{eqnarray}
where
\begin{eqnarray*}
& & \sup_{0 \leq t \leq \nu^*} |\epsilon_0(t)|    \\
&=& \sup_{0 \leq t \leq \nu^*} \Big| n^{-1/2} \sum_{i=1}^n \int_t^{\nu^*} \Big\{ \frac{ 1 }{ E\{S_C(s|\gamma_i) \gamma_i\} \mu_0(s) }
           -  \frac{ 1 }{ n^{-1} \sum_{i=1}^n I( C_i \geq s ) N_i(s) }  \Big\}   d M_i(s) \Big|       \\
& \leq & \sup_{0 \leq t \leq \nu^*} \left| \frac{ 1 }{ E\{S_C(t|\gamma_i) \gamma_i\} \mu_0(t) }
           -  \frac{ 1 }{ n^{-1} \sum_{i=1}^n I( C_i \geq t ) N_i(t) } \right| \int_0^{\nu^*} \Big| n^{-1/2} \sum_{i=1}^n d M_i(s)  \Big|   \\
&=& o_p(1).
\end{eqnarray*}
{
Because $\varphi_i(t) (i=1, \ldots, n)$ can be written as the summation of monotone functions of $t$, $\{ \varphi_i(t): t \in [0, \nu^*] \}$ is a Donsker class \citep{van1996}. By the Donsker theorem, $n^{1/2} \{ \hat{H}(t) - H_0(t) \}$ converges weakly to a tight Gaussian process with mean zero and covariance function $E[ \varphi_i(s) \varphi_i(t) ]$ at $(s, t)$.
}

By the continuity of the function $\exp(x)$, $\hat{\mu}(t)$ is a uniformly consistent estimator of $\mu_0(t)$.
Taylor expansions of $\exp\{ H(t) \}$ around $H(t)=H_0(t)$, along with the fact that $\hat{H}(t)$ uniformly converges to $H_0(t)$, gives that
\begin{eqnarray}
  & &  n^{1/2} \{ \hat{\mu}(t) - \mu_0(t) \}       \nonumber      \\
  &=&  n^{1/2} \{ \exp\{ \hat{H}(t) \} - \exp\{ H_0(t) \} \}        \nonumber      \\
  &=&  n^{1/2} \exp\{ H_0(t) \} \{ \hat{H}(t) - H_0(t) \} + \epsilon(t)         \nonumber      \\
  &=&  - n^{-1/2} \sum_{i=1}^n \mu_0(t) \varphi_i(t) + \mu_0(t) \epsilon_0(t) + \epsilon(t),               \label{eqmu}
\end{eqnarray}
where $\sup_t | \mu_0(t) \epsilon_0(t) + \epsilon(t) | \rightarrow 0$, the last equality follows from (\ref{eqh}).
Hence $n^{1/2} \{ \hat{\mu}(t) - \mu_0(t) \}$ converges weakly to a zero-mean Gaussian process with the covariance function $ E[ \mu_0(s) \psi_i(s) \psi_i(t) \mu_0(t)]$ at $(s, t)$.

\setcounter{theorem}{0}
\begin{corollary} \label{cor1}
Under the regularity condition C1(a), 
$$\hat{\mu}(C) - \mu_0(C) = - n^{-1} \sum_{i=1}^n \mu_0(C) \varphi_i(C) + o_p(1).$$
\end{corollary}
Corollary \ref{cor1} follows immediately from (\ref{eqmu}).

\vspace{5mm}

{ In the following, we present Lemmas \ref{lem1} and \ref{lem2}, which are needed for the proof of Theorems \ref{theo1} and \ref{theo2}.}
Define $F_{\bbeta}(\gamma) = \inf\{\tau: \exp\{\bX^{\T} \bbeta(\tau)\} > \gamma \}$ as the inverse function of $\exp\{\bX^{\T} \bbeta(\tau)\}$.
For a vector $\bu$, we use $| \bu |$ to denote its componentwise absolute values. $|\bu| < 1$ means that each component of $\bu$ is bounded by $-1$ and $1$.

Note that we can only obtain the sequence $\{ \hat{\bbeta}(\tau_k)\}_{k=1}^K$ from the estimation algorithm introduced in subsection \ref{ssea}. This sequence determines the whole quantile function $\hat{\bbeta}(\tau)$, the conditional density $g\{\gamma|\bX; \hat{\bbeta}(\cdot) \}$, and further $f\{ \gamma|m, C, \bX; \hat{\bbeta}(\cdot), \hat{\mu}(\cdot) \}$. In the following Lemma \ref{lem1}, we establish that the distance between two density functions is negligible as long as the corresponding sequences are close enough.

\begin{lemma}   \label{lem1}
Suppose the regularity condition C1 holds. For any $\mu \in \mathcal{U}$, two sequences $\{ \tilde{\bbeta}_1(\tau_k)\}_{k=1}^K$ and $\{ \tilde{\bbeta}_2(\tau_k)\}_{k=1}^K$  that satisfy $\max_{1 \leq k \leq K}|\tilde{\bbeta}_1(\tau_k)-\tilde{\bbeta}_2(\tau_k) |=o(1)$, we have
$$
\sup_{ \gamma > 0 } | f \{ \gamma|m, C, \bX; \tilde{\bbeta}_1(\cdot), \mu(\cdot) \} - f \{ \gamma|m, C, \bX; \tilde{\bbeta}_2(\cdot), \mu(\cdot) \} | = o(1),
$$
where $\tilde{\bbeta}_i(\cdot)$ is the piecewise linear function induced by the sequence $\{ \tilde{\bbeta}_i(\tau_k)\}_{k=1}^K$, $i=1,2$.
\end{lemma}

{\it Proof of Lemma \ref{lem1}.}
Due to the boundedness of $\rho\{ m |\gamma, C; \mu(\cdot)\}$ imposed in condition C1(c), it suffices to show that
$$
\sup_{ \gamma > 0 } | g \{\gamma|\bX; \tilde{\bbeta}_1(\cdot) \} - g \{ \gamma|\bX;  \tilde{\bbeta}_2(\cdot) \} | = o(1),
$$
where $g \{\gamma|\bX; \tilde{\bbeta}_i(\cdot) \}$ is the density function corresponding to the quantile function $\exp\{\bX^{\T} \tilde{\bbeta}_i(\tau)\}$, $i=1, 2$.
Denote $L = \min\{ \exp \{\bX^{\T} \tilde{\bbeta}_1(\tau_1) \}, \exp \{\bX^{\T} \tilde{\bbeta}_2(\tau_1) \} \}$ and  $U=\max\{ \exp \{\bX^{\T} \tilde{\bbeta}_1(\tau_K) \}, \\
\exp \{\bX^{\T} \tilde{\bbeta}_2(\tau_K) \}$, then we have $g\{\gamma|\bX; \tilde{\bbeta}_1(\cdot)\} = g\{\gamma|\bX; 
 \tilde{\bbeta}_2(\cdot)\} = 0$ when $\gamma < L$ or $\gamma > U$.

Next, we show $\sup_{\gamma \in [L, U]} |g\{\gamma|\bX; \tilde{\bbeta}_1(\cdot)\} - g\{\gamma|\bX; \tilde{\bbeta}_2(\cdot)\}| = o(1) $.
The piecewise linear function $\tilde{\bbeta}_i(\cdot)$ induced by the sequence $\{ \tilde{\bbeta}_i(\tau_k)\}_{k=1}^K$ has the form
\begin{eqnarray*}
\tilde{\bbeta}_i(\tau) = \left\{
\begin{array}{ll}
  \tilde{\bbeta}_i( \tau_1 ), & ~~~~ \tau \leq \tau_1, \\
  \tilde{\bbeta}_i( \tau_k )  + \frac{ \tau - \tau_k }{ \tau_{k+1} - \tau_k } \left\{ \tilde{\bbeta}_i( \tau_{k+1} ) - \tilde{\bbeta}_i(\tau_k)  \right\}, & ~~~~\tau_k < \tau \leq \tau_{k+1},     \\
 \tilde{\bbeta}_i( \tau_K ), & ~~~~ \tau > \tau_K,
\end{array}
 \right.
\end{eqnarray*}
$k=1, \ldots, K-1$ for $i=1, 2$.
The difference between $ \tilde{\bbeta}_1(\tau) $ and $ \tilde{\bbeta}_2(\tau) $ is bounded by
\begin{eqnarray*}
&  &  | \tilde{\bbeta}_1(\tau) - \tilde{\bbeta}_2(\tau) |   \\
& \leq &  \left\{
\begin{array}{ll}
 | \tilde{\bbeta}_1(\tau_1) - \tilde{\bbeta}_2(\tau_1) |, & ~~~~ \tau \leq \tau_1, \\
 2~| \tilde{\bbeta}_1(\tau_k) - \tilde{\bbeta}_2(\tau_k) |  + | \tilde{\bbeta}_1(\tau_{k+1}) - \tilde{\bbeta}_2(\tau_{k+1}) |, & ~~~~\tau_k < \tau \leq \tau_{k+1},   \\
 | \tilde{\bbeta}_1(\tau_K) - \tilde{\bbeta}_2(\tau_K) |, & ~~~~ \tau > \tau_K{,}
\end{array}
 \right.
\end{eqnarray*}
$k=1, \ldots, K-1$.
Since $\max_{1 \leq k \leq K}|\tilde{\bbeta}_1(\tau_k)-\tilde{\bbeta}_2(\tau_k) |=o(1)$, the boundedness of $\bX$ imposed in condition C1(b), and the continuity of function $\exp(x)$, we have $\sup_{\tau} |\exp\{\bX^{\T} \tilde{\bbeta}_1(\tau)\} - \exp\{\bX^{\T} \tilde{\bbeta}_2(\tau)\} | = o(1)$.
{
This, coupled with the fact
\begin{eqnarray*}
g\{\gamma|\bX; \tilde{\bbeta}_i(\cdot)\}
&=& \sum_{k=1}^K \frac{ \tau_{k} - \tau_{k-1} }{ \exp\{ \bX^{\T} \tilde{\bbeta}_i ( \tau_{k} ) \} -
        \exp \{ \bX^{\T} \tilde{\bbeta}_i(\tau_{k-1}) \} }   \\
& & ~~~~~~~~ \cdot I \{ \exp \{ \bX^{\T} \tilde{\bbeta}_i(\tau_{k-1}) \} < \gamma \leq \exp\{ \bX^{\T} \tilde{\bbeta}_i ( \tau_{k} ) \} \}, i=1,2,
\end{eqnarray*}
implies that $\sup_{\gamma \in [L, U]} |g\{\gamma|\bX; \tilde{\bbeta}_1(\cdot)\} - g\{\gamma|\bX; \tilde{\bbeta}_2(\cdot)\}| = o(1) $.
Hence we complete the proof of Lemma \ref{lem1}.
%
}

\begin{lemma}    \label{lem2}
Under the regularity conditions C1--C4, if $K \rightarrow \infty$ and $K/n^{\alpha} \rightarrow 0$ for some $\alpha > 0$, then for any $\mu \in \mathcal{U}$, we have
$$
 \sup_{1 \leq k \leq K}  \| \bS_n( \hat{\bbeta}, \mu, \tau_k) - \bs ( \hat{\bbeta}, \mu, \tau_k) \|  = o_p(1) ~~~~~\mbox{as} ~~~~n \rightarrow \infty.
$$
\end{lemma}

{\it Proof of Lemma \ref{lem2}.} Define $ \mathcal{G} = \{ \bbeta:[\tau_1, \tau_K] \rightarrow \mathbb{R}^p, \bbeta(\cdot)$ is a piecewise linear function whose knots are $ \mathcal{S}_K \}$.
For any $\epsilon > 0$, it is sufficient to show that
$$
\Pr \left( \sup_{ \bbeta \in \mathcal{G} } \sup_{ 1 \leq k \leq K }  \| \bS_n( \bbeta, \mu, \tau_k) - \bs ( \bbeta, \mu, \tau_k) \|  > \epsilon \right) \rightarrow 0
$$
as $n \rightarrow \infty$.
Following \cite{Wei2009}, we use Huber's chaining argument to show it.
Without loss of generality, {we assume $ \mathcal{G} = \{ \bbeta: \bbeta \in \mathcal{G},~ \sup_{1 \leq k \leq K} |\bbeta(\tau_k) - \bbeta_0(\tau_k)| < 1 \}$.
As stated before, $ \mathcal{G} $ is determined by a parameter space $\{ ( \bbeta(\tau_1), \ldots, \bbeta(\tau_K) ), 
 \bbeta \in \mathcal{G} \}$, which can be partitioned into $L_n$ disjoint small cubes $\Gamma_l$ with diameters less than $q_n = C_g K/n^{\alpha} = o(1)$, where $C_g$ is a constant.
Let $\bbeta_l(\cdot)$ be the piecewise linear function induced by the center of the $l$th cube $\Gamma_l$.}
Note that
\begin{eqnarray*}
& & \Pr \Big( \sup_{ \bbeta \in \mathcal{G} } \sup_{1 \leq k \leq K} \| \bS_n(\bbeta, \mu, \tau_k) - \bs(\bbeta, \mu, \tau_k)  \| > \epsilon \Big)  \\
& \leq & \Pr \Big( \max_{ 1 \leq l \leq L_n }  \sup_{ \bbeta \in \Gamma_l  } \sup_{1 \leq k \leq K} \| \bS_n(\bbeta, \mu, \tau_k) - \bS_n(\bbeta_l, \mu, \tau_k)  \\
& & ~~~~~~~ - \bs(\bbeta, \mu, \tau_k) + \bs(\bbeta_l, \mu, \tau_k)  \| > \epsilon/2 \Big)    \\[2mm]
& & +  \Pr \Big( \max_{ 1 \leq l \leq L_n } \sup_{1 \leq k \leq K} \| \bS_n(\bbeta_l, \mu, \tau_k)  - \bs(\bbeta_l, \mu, \tau_k)  \| > \epsilon/2 \Big)    \\
& \doteq & P_1 + P_2,
\end{eqnarray*}
where $\bbeta \in \Gamma_l$ means the parameter $\{ \bbeta(\tau_k) \}_{k=1}^K \in \Gamma_l$.
In the following, we show $P_1=o(1)$ and $P_2=o(1)$.

First, we have
\begin{eqnarray*}
 & &  \| \bS_n(\bbeta, \mu, \tau_k) - \bS_n(\bbeta_l, \mu, \tau_k)  \|     \\
 & \leq & \bigg\| \frac{1}{n} \sum_{i=1}^n \int_{\gamma} \left[ \psi_{\tau_k} \{ \log(\gamma) - \bX_i^{\T}\bbeta(\tau_k) \} -  \psi_{\tau_k} \{ \log(\gamma) - \bX_i^{\T}\bbeta_l(\tau_k) \} \right] \cdot \bX_i    \\
 & & ~~~~~~~~~~~~~~~~~~  \cdot f\{\gamma|m_i, C_i, \bX_i; \bbeta_l( \cdot ), \mu(\cdot) \} d \gamma   \bigg\|     \\
 & & +  \bigg\| \frac{1}{n} \sum_{i=1}^n \int_{\gamma} \psi_{\tau_k} \{ \log(\gamma) - \bX_i^{\T}\bbeta(\tau_k) \}  \cdot \bX_i    \\
 & & ~~~~~~~~~~~~~~~~~~  \cdot \left[ f\{ \gamma|m_i, C_i, \bX_i; \bbeta(\cdot), \mu(\cdot) \} - f\{\gamma|m_i, C_i, \bX_i; \bbeta_l(\cdot), \mu(\cdot) \} \right] d \gamma   \bigg\|     \\
 & \doteq & \mbox{S}_1 + \mbox{S}_2.
\end{eqnarray*}
Under the regularity condition C1(b), there exists a constant $C_1$ such that
\begin{eqnarray*}
& & \max_{1 \leq l \leq L_n}  \sup_{\bbeta \in \Gamma_l} \sup_{ 1 \leq k \leq K } \mbox{S}_1     \\
& \leq & \max_{1 \leq l \leq L_n}  \sup_{\bbeta \in \Gamma_l} \sup_{ 1 \leq k \leq K } \Bigg\| \frac{1}{n} \sum_{i=1}^n \int_{\gamma} I\{ |\bX_i^{\T} \bbeta_l(\tau_k) - \log(\gamma)| \leq |\bX_i^{\T}\{\bbeta_l(\tau_k)-\bbeta(\tau_k)\}| \}
                        \\
& &~~~~~~~~~~~~~~~~\cdot \bX_i  \cdot f\{\gamma|m_i, C_i, \bX_i; \bbeta_l(\cdot), \mu(\cdot) \} d \gamma   \Bigg\|     \\
& \leq & \max_{1 \leq l \leq L_n} \sup_{ 1 \leq k \leq K }  \frac{1}{n} \sum_{i=1}^n \int_{\gamma} I\{ |\bX_i^{\T}\bbeta_l(\tau_k) - \log(\gamma)| \leq \| \bX_i \| q_n \}
                  \cdot  \| \bX_i \|    \\
& &  ~~~~~~~~~~~~~~~~      \cdot f\{\gamma|m_i, C_i, \bX_i; \bbeta_l(\cdot), \mu(\cdot)\} d \gamma     \\
& \leq & C_1  \cdot \max_{1 \leq l \leq L_n} \sup_{ 1 \leq k \leq K }  \frac{1}{n} \sum_{i=1}^n \int_{\gamma} I\{ |\bX_i^{\T}\bbeta_l(\tau_k) - \log(\gamma)| \leq C_1 q_n \}  \\
& &   ~~~~~~~~~~~~~~~~   \cdot f\{\gamma|m_i, C_i, \bX_i; \bbeta_l(\cdot), \mu(\cdot)\} d \gamma      \\
& = & C_1  \cdot \max_{1 \leq l \leq L_n} \sup_{ 1 \leq k \leq K }  \frac{1}{n} \sum_{i=1}^n \Pr\{ |\bX_i^{\T}\bbeta_l(\tau_k) - \log(\gamma)| \leq C_1 q_n | m_i, C_i, \bX_i; \bbeta_l(\cdot), \mu(\cdot) \}.
\end{eqnarray*}
Let $g_i(z)$ be the density of $\bX_i^{\T}\bbeta_l(\tau_k) - \log(\gamma)$ given $\{m_i, C_i, \bX_i, \bbeta_l(\cdot), \mu(\cdot)\}$. Then following the regularity condition C1, $g_i(z)$ is also continuous and bounded away from zero and infinity. Following the mean value theorem, for any $i$ there exists $z_i^*$ such that $\Pr\{ |\bX_i^{\T}\bbeta_l(\tau_k) - \log(\gamma)| \leq C_1 q_n | m_i, C_i, \bX_i; \bbeta_l(\cdot), \mu(\cdot) \} = 2 C_1 q_n g_i(z_i^*)$. It follows that $ \max_{1 \leq l \leq L_n}  \sup_{\bbeta \in \Gamma_l} \sup_{1 \leq k \leq K} \mbox{S}_1 =O_p(q_n)=o_p(1)$.
{ Regarding $S_2$}, a sufficient condition for $\max_{1 \leq l \leq L_n}  \sup_{\bbeta \in \Gamma_l} \sup_{1 \leq k \leq K} \mbox{S}_2 = o_p(1)$ is that $\sup_{ \gamma } |f\{\gamma|m_i, C_i, \bX_i; 
\bbeta(\cdot), \mu(\cdot)\} - f\{\gamma|m_i, C_i, \bX_i; \bbeta_l(\cdot), \mu(\cdot)\}| = o_p(1)$ for any $1 \leq l \leq L_n$, $\bbeta \in \Gamma_l$ and $1 \leq k \leq K$. This immediately holds according to Lemma \ref{lem1}. Following a similar argument, we can also show that $\max_{1 \leq l \leq L_n}  \sup_{\bbeta \in \Gamma_l} \sup_{1 \leq k \leq K} 
\| \bs(\bbeta, \mu, \tau_k)-\bs(\bbeta_l, \mu, \tau_k) \| = o_p(1)$. It then follows that $P_1=o(1)$.

To show $P_2=o(1)$, a sufficient condition  is that
$$
\Pr \left( \max_{1 \leq l \leq L_n, 1 \leq k \leq K, 1 \leq m \leq p}  \frac{1}{n} \left| \sum_{i=1}^n
    \left[\zeta_i(l, k, m)-E\{\zeta_i(l, k, m)\}  \right]   \right| > \epsilon/2 \right) = o(1),
$$
where $\zeta_i(l, k, m) = \int_\gamma \psi_{\tau_k} \{ \log(\gamma)-\bX_i^{\T} \bbeta_l(\tau_k) \}\cdot \bX_i[m] \cdot f\{\gamma|m_i, C_i, \bX_i; \bbeta_l(\cdot), \mu(\cdot) \} d \gamma $ with $\bX_i[m]$ as the $m$th component of $\bX_i$.
Under the regularity condition C1, there exists a constant $C_2$ such that $| \zeta_i(l, k, m) | < C_2$ for every $i$.
{Hence, we have}
\begin{eqnarray*}
& & \Pr \left( \max_{1 \leq l \leq L_n, 1 \leq k \leq K, 1 \leq m \leq p}  \frac{1}{n} \left| \sum_{i=1}^n \left[\zeta_i(l, k, m)-E\{\zeta_i(l,k, m)\}  \right]   \right| > \epsilon/2 \right) \\
& \leq &  L_n \cdot K \cdot p \cdot \Pr \left(  \frac{1}{n} \left| \sum_{i=1}^n \left[\zeta_i(l,k,m)-E\{\zeta_i(l,k,m)\}  \right]   \right| > \epsilon/2 \right)   \\
& \leq & 2 \cdot L_n \cdot K \cdot p \cdot \exp \left( -\frac{3 n^2 \epsilon^2 }{ 24 n C_2^2 + 4 n C_2 \epsilon } \right) = o(1),
\end{eqnarray*}
{ where the last inequality follows from the Bernstein's inequality.}
Hence we complete the proof of Lemma \ref{lem2}.

{\it Proof of Theorem \ref{theo1}.}
First, we show $\sup_{\tau \in [\tau_1, \tau_K]} \| \bs( \hat{\bbeta}, \mu_0, \tau ) - \bs ( \bbeta_0, \mu_0, \tau ) \| \stackrel{p}{\longrightarrow} 0$.
Note that
\begin{eqnarray*}
& &  \sup_{\tau \in [\tau_1, \tau_K]} \| \bs( \hat{\bbeta}, \mu_0, \tau ) - \bs ( \bbeta_0, \mu_0, \tau ) \|      \\
& \leq & \sup_{ 1 \leq k \leq K } \| \bS_n( \hat{\bbeta}, \hat{\mu}, \tau_k ) - \bs( \hat{\bbeta}, \hat{\mu}, \tau_k ) \|
  + \sup_{ 1 \leq k \leq K } \sup_{\tau \in [\tau_k, \tau_{k+1})} \| \bs( \hat{\bbeta}, \hat{\mu}, \tau_k ) - \bs( \hat{\bbeta}, \hat{\mu}, \tau ) \|      \\
& &  + \sup_{\tau \in [\tau_1, \tau_K]} \| \bs( \hat{\bbeta}, \hat{\mu}, \tau ) - \bs( \hat{\bbeta}, \mu_0, \tau ) \| + \sup_{ 1 \leq k \leq K } \| \bS_n( \hat{\bbeta}, \hat{\mu}, \tau_k ) \|  \\
& &  + \sup_{\tau \in [\tau_1, \tau_K]} \| \bs ( \bbeta_0, \mu_0, \tau ) \|  \\
& \doteq & \mbox{I}_1 + \mbox{I}_2 + \mbox{I}_3 + \mbox{I}_4 + \mbox{I}_5.
\end{eqnarray*}
For $\mbox{I}_1$, we can obtain from Lemma \ref{lem2} that $\sup_{ 1 \leq k \leq K } \| \bS_n( \hat{\bbeta}, \hat{\mu}, \tau_k ) - \bs( \hat{\bbeta}, \hat{\mu}, \tau_k )\| = o_p(1)$.
For $\mbox{I}_2$, under the regularity condition C1, there exists a constant $C_3 > 0$ such that
\begin{eqnarray*}
& & \sup_{ 1 \leq k \leq K } \sup_{\tau \in [\tau_k, \tau_{k+1})} \| \bs( \hat{\bbeta}, \hat{\mu}, \tau_k ) - \bs( \hat{\bbeta}, \hat{\mu}, \tau ) \| \leq C_3  \sup_{ 1 \leq k \leq K } \sup_{\tau \in [\tau_k, \tau_{k+1})}| \tau - \tau_k |  \\
& \leq &  C_3  \sup_{ 1 \leq k \leq K }  | \tau_{k+1} - \tau_k | \leq C_3 \| \mathcal{S}_K \| \rightarrow 0.
\end{eqnarray*}
To prove $\mbox{I}_3=o_p(1)$, note that $\bs(\bbeta, \mu, \tau)$ depends on $\mu(\cdot)$ only through $\mu(C)$, the functional derivative of $\bs(\bbeta, \mu, \tau)$ with respect to $\mu$ is bounded by the regularity condition C1.
This, coupled with the uniform consistency of $\hat{\mu}(\cdot)$ shown in Lemma \ref{lem0}, implies that $\sup_{\tau \in [\tau_1, \tau_K]} \| \bs( \hat{\bbeta}, \hat{\mu}, \tau ) - \bs( \hat{\bbeta}, \mu_0, \tau ) \| = o_p(1)$.
For $\mbox{I}_4$, according to the definition of $\hat{\bbeta}(\cdot)$, we have $ \sup_{ 1 \leq k \leq K } \| \bS_n( \hat{\bbeta}, \hat{\mu}, \tau_k ) \| = o(1)$.
For $\mbox{I}_5$, the multiplicative intensity model (\ref{model1}), coupled with the quantile regression model (\ref{model2}), implies that $\sup_{\tau \in [\tau_1, \tau_K]} \| \bs ( \bbeta_0, \mu_0, \tau ) \|=0$.
Hence we have proven
\begin{eqnarray}  \label{eqa1}
\bs( \hat{\bbeta}, \mu_0, \tau ) - \bs ( \bbeta_0, \mu_0, \tau ) = - [ \bupsilon( \hat{\bbeta} ) - \bupsilon( \bbeta_0 ) ] = \bo_{[\tau_1, \tau_K]}(1),
\end{eqnarray}
where $\bo_I(1)$ denotes a term that converges to 0 in probability uniformly on the interval $I$.
{ Then (\ref{eqa1})} implies
\begin{eqnarray}  \label{eqa2}
 \bupsilon( \hat{\bbeta} ) - \bupsilon ( \bbeta_0  ) = \bo_{[\tau_1, \tau_K]}(1).
\end{eqnarray}
Note that $\bupsilon(\bbeta)$ is a functional of $\bbeta$. By condition C3(a), $\bupsilon$ is Fr$\acute{\mbox{e}}$chet differentiable at $\bbeta_0$. Hence, for any direction $\bh \in \mathcal{F}$ and $\bbeta_0 + t \bh \in \mathcal{D}$, there is a linear map $\dot{\bupsilon}_{\bbeta_0}$ such that
$$
\frac{ \bupsilon( \bbeta_0 + t \bh ) - \bupsilon( \bbeta_0 ) }{t}   \rightarrow   \dot{\bupsilon}_{\bbeta_0}( \bh ) ~~~\mbox{as}~~~ t \rightarrow 0.
$$
Let $\bh = (\hat{\bbeta} - \bbeta_0)/t$, we have
\begin{eqnarray}  \label{eqa3}
\{ \bupsilon( \hat{\bbeta} ) - \bupsilon( \bbeta_0 ) \} - t \dot{\bupsilon}_{\bbeta_0}\{ (\hat{\bbeta}-\bbeta_0)/t \}  \rightarrow  0 ~~~\mbox{as}~~~ t \rightarrow 0.
\end{eqnarray}
By (\ref{eqa2}), (\ref{eqa3}), and the linearity of $\dot{\bupsilon}_{\bbeta_0}$, we immediately have
\begin{eqnarray}  \label{eqa4}
 \dot{\bupsilon}_{\bbeta_0}( \hat{\bbeta}-\bbeta_0 )    = \bo_{[\tau_1, \tau_K]}(1).
\end{eqnarray}
Since $\hat{\bbeta}$ and $\bbeta_0$ are continuous on $[\tau_1, \tau_K]$, by condition C3(b), (\ref{eqa4}) implies
$$
\sup_{\tau \in [\tau_1, \tau_K]} \| \hat{\bbeta}(\tau) - \bbeta_0(\tau) \| = \bo_p(1).
$$
This completes the proof of Theorem \ref{theo1}.

\begin{lemma}   \label{lem3}
Denote $\check{\bbeta}_0(\tau)$ as the piecewise linear function induced by the sequence $\{ \bbeta_0(\tau_k) \}_{k=1}^K$. Under the regularity conditions C1--C4, if $\lim_{n \rightarrow \infty} \| \mathcal{S}_K \| = 0$, then for any $\mu \in \mathcal{U}$, we have
\begin{eqnarray*}
& & \sup_{ \gamma > 0 }| f\{ \gamma|m, C, \bX; \check{\bbeta}_0(\cdot), \mu(\cdot) \} - f\{ \gamma| m, C, \bX; \bbeta_0(\cdot), \mu(\cdot) \}|   \\
& & ~~~~ \cdot I\{ \bX^{\T} \bbeta_0(\tau_1) < \log(\gamma) \leq \bX^{\T} \bbeta_0(\tau_K)   \} = o(1).
\end{eqnarray*}
\end{lemma}

{\it Proof of Lemma \ref{lem3}.}
Recall that
$$
f \{\gamma | m, C, \bX; \bbeta(\cdot){, \mu(\cdot)} \} = \frac{ \rho\{ m | \gamma, C; \mu(\cdot) \}  g\{\gamma | \bX; \bbeta(\cdot)\} } { \int_\gamma  \rho\{ m | \gamma, C; \mu(\cdot) \}  g\{\gamma | \bX; \bbeta(\cdot)\} d \gamma}.
$$
According to the boundedness of $\rho\{ m | \gamma, C; \mu(\cdot) \}$ imposed in condition C1(c), it is sufficient to show that
\begin{eqnarray*}
\sup_{ \gamma > 0 } \left|  g\{\gamma | \bX; \check{\bbeta}_0(\cdot)\} - g\{\gamma | \bX; \bbeta_0(\cdot)\} \right| \cdot I\{ \bX^{\T} \bbeta_0(\tau_1) < \log(\gamma) \leq \bX^{\T} \bbeta_0(\tau_K)   \} = o(1)
\end{eqnarray*}
holds for any $\bX$.

Let $F_{\bX}(\gamma) = \inf\{\tau: \exp\{ \bX^{\T} \bbeta_0(\tau) \} \geq \gamma \}$ be the quantile rank of $\gamma$ with respect to the probability measure induced by the quantile function $\exp\{ \bX^{\T} \bbeta_0(\tau) \} $, and let $h_{\bX}(\tau) = 1/(\exp\{ \bX^{\T} \bbeta_0(\tau) \})^\prime$ be the density function of $\gamma$ at the $\tau$th quantile.
For any $\gamma$ that is bounded between $\exp\{ \bX^{\T} \bbeta_0(\tau_1)\}$ and $\exp\{ \bX^{\T} \bbeta_0(\tau_K)\}$, there exists a $k$ such that $ \bX^{\T} \bbeta_0(\tau_{k}) < \log(\gamma) \leq \bX^{\T} \bbeta_0(\tau_{k+1}) $. Consequently,
{
\begin{eqnarray*}
& &  \sup_{\gamma > 0} \left|  g\{\gamma | \bX; \check{\bbeta}_0(\cdot)\} - g\{\gamma | \bX; \bbeta_0(\cdot)\} \right| \cdot I\{ \bX^{\T} \bbeta_0(\tau_1) < \log(\gamma) \leq \bX^{\T} \bbeta_0(\tau_K)   \}   \\
& = & \sup_{\substack{ 1 \leq k \leq K \\ \bX^{\T} \bbeta_0(\tau_k) < \log(\gamma) \leq \bX^{\T} \bbeta_0(\tau_{k+1})}}   \Big|  \frac{\tau_{k+1} - \tau_{k}}{\exp\{\bX^{\T} \bbeta_0(\tau_{k+1})\} -\exp\{\bX^{\T} \bbeta_0(\tau_{k})\}}   \\
& & ~~~~~~~~~~~~~ - \frac{1}{[\exp\{ \bX^{\T} \bbeta_0\{ F_\bX(\gamma) \} \}]^\prime}  \Big|  \\  [2mm]
& = & \sup_{\substack{ 1 \leq k \leq K \\ \bX^{\T} \bbeta_0(\tau_k) < \tau_*, ~\log(\gamma)  \leq \bX^{\T} \bbeta_0(\tau_{k+1})}}  \left| h_{\bX}(\tau_*) - h_{\bX}\{ F_\bX(\gamma) \} \right|      \\  [2mm]
& \leq & \sup_{\substack{ 1 \leq k \leq K \\ \bX^{\T} \bbeta_0(\tau_k) < \tau_{**} \leq \bX^{\T} \bbeta_0(\tau_{k+1})}}  \left| h_{\bX}^\prime( \tau_{**} ) \right| | \tau_{k+1} - \tau_{k} |,
\end{eqnarray*}
}
where $\tau_{**}$ is between $\tau_*$ and $F_\bX(\gamma)$.
According to the boundedness condition of $h_{\bX}^\prime( \tau )$ implied in the regularity condition C4(b) and the assumption $\lim_{n \rightarrow \infty} \| \mathcal{S}_K \| = 0$, we can complete the proof of Lemma \ref{lem3}.

\begin{lemma}  \label{lem4}
Suppose the regularity conditions C1--C4 hold.
For any piecewise linear function $\tilde{\bbeta}(\cdot)  \in \mathcal{G}$ that satisfies $\sup_{\tau \in [\tau_1, \tau_K]} \| \tilde{\bbeta}(\tau) - \bbeta_0(\tau) \| \stackrel{p}{\longrightarrow} 0$,
and any uniformly consistent estimator $\tilde{\mu}(\cdot) \in \mathcal{U}$ that satisfies $\sup_{t \in [0, \nu^*]} \| \tilde{\mu}(t) - \mu_0(t) \| \stackrel{p}{\longrightarrow} 0$,
 we have
\begin{eqnarray*}
\sup_{\tau \in [\tau_1, \tau_K]}  \left\| n^{1/2} \{ \bS_n( \tilde{\bbeta}, \tilde{\mu}, \tau ) - \bS_n(\bbeta_0, \mu_0, \tau )\} - n^{1/2} \{ \bs( \tilde{\bbeta}, \tilde{\mu}, \tau ) - \bs(\bbeta_0, \mu_0, \tau ) \}  \right\|  \stackrel{p}{\longrightarrow} 0.
\end{eqnarray*}
\end{lemma}

{\it Proof of Lemma \ref{lem4}.}
Denote $\xi( \bbeta, \mu, \tau ) = \int_{\gamma} \psi_{\tau} \{ \log(\gamma) - \bX^{\T} \bbeta(\tau) \} f\{ \gamma | m, C, \bX; \bbeta( \cdot ),  \\
 \mu(\cdot) \}    d \gamma $ for notation simplicity.
Further define $ s_1( \bbeta, \mu, \tau )= E [ \xi ( \bbeta, \mu,  \tau ) ]$ and   
$ \sigma_d^2( \tilde{\bbeta}, \tilde{\mu}, \tau)  = \mbox{Var}[ \xi( \tilde{\bbeta}, \tilde{\mu}, \tau ) - \xi( \bbeta_0, \mu_0, \tau ) - s_1( \tilde{\bbeta}, \tilde{\mu}, \tau )  + s_1(\bbeta_0, \mu_0, \tau )]$.
According to \cite{Alexander1984} and \cite{Lai1988}, it is sufficient to show that $\sigma_d^2( \tilde{\bbeta}, \tilde{\mu}, \tau) \stackrel{p}{\longrightarrow} 0$, given the boundedness of $\bX$.

Note that
\begin{eqnarray*}
\sigma_d^2( \tilde{\bbeta}, \tilde{\mu}, \tau)
& \leq & E \Big\{  \int_{\gamma} \Big[ \psi_{\tau} \{ \log(\gamma) - \bX^{\T} \tilde{\bbeta}(\tau) \} f\{ \gamma | m, C, \bX; \tilde{\bbeta}( \cdot ), \tilde{\mu}(\cdot) \}  \\
& & ~~~~~~~~ - \psi_{\tau} \{ \log(\gamma) - \bX^{\T} \bbeta_0(\tau) \} f\{ \gamma |m, C, \bX; \bbeta_0( \cdot ), \mu_0(\cdot) \} \Big]   d \gamma \Big\}^2  \\
& \leq & 2 \cdot E \Big\{  \int_{\gamma}  \psi_{\tau} \{ \log(\gamma) - \bX^{\T} \tilde{\bbeta}(\tau) \} \big[ f\{ \gamma | m, C, \bX; \tilde{\bbeta}( \cdot ), \tilde{\mu}(\cdot) \}    \\
& & ~~~~~~~~ -  f\{ \gamma | m, C, \bX; \tilde{\bbeta}( \cdot ), \mu_0(\cdot) \} \big]   d \gamma \Big\}^2  \\
&& + 2 \cdot E \Big\{  \int_{\gamma} \left[ \psi_{\tau} \{ \log(\gamma) - \bX^{\T} \tilde{\bbeta}(\tau) \} - \psi_{\tau} \{ \log( \gamma ) - \bX^{\T} \bbeta_0(\tau) \} \right] \\
& & ~~~~~~~~ \cdot f\{ \gamma | m, C, \bX; \tilde{\bbeta}( \cdot ), \mu_0(\cdot) \}   d \gamma \Big\}^2  \\
&& + 2 \cdot E \Big\{  \int_{\gamma}  \psi_{\tau} \{ \log(\gamma) - \bX^{\T} \bbeta_0(\tau) \} \big[ f\{ \gamma | m, C, \bX; \tilde{\bbeta}( \cdot ), \mu_0(\cdot) \}  - \\
& & ~~~~~~~~ f\{ \gamma | m, C, \bX; \bbeta_0( \cdot ), \mu_0(\cdot) \} \big]   d \gamma \Big\}^2  \\
& \doteq & \mbox{II}_1 + \mbox{II}_2 + \mbox{II}_3.
\end{eqnarray*}
$\mbox{II}_1 = o_p(1)$ follows from the fact that the functional derivative of $f\{ \gamma | m, C, \bX; \tilde{\bbeta}( \cdot ), \mu(\cdot) \}$ at $\mu_0$ in the direction $[\tilde{\mu}-\mu_0]$ is bounded and the uniform consistency of $\tilde{\mu}(\cdot)$.
Next, we show $\mbox{II}_2 = o_p(1)$ and $\mbox{II}_3 = o_p(1)$.
For $\mbox{II}_2$, notice that
\begin{eqnarray*}
 \mbox{II}_2 & = & E \Big\{  \int_{\gamma} \left[ I \{ \log(\gamma) - \bX^{\T} \tilde{\bbeta}(\tau) < 0 \} - I \{ \log(\gamma) - \bX^{\T} \bbeta_0(\tau) < 0 \} \right]^2  \\
 & & ~~~~~ \cdot  f\{ \gamma | m, C, \bX; \tilde{\bbeta}( \cdot ), \mu_0(\cdot) \}   d \gamma \Big\}   \\
    & = & E \Big\{  E \big[ \left(I \{ \log(\gamma) - \bX^{\T} \tilde{\bbeta}(\tau) < 0 \} - I \{ \log(\gamma) - \bX^{\T} \bbeta_0(\tau) < 0 \} \right)^2 \left|\right.  \\
    & & ~~~~~ m, C, \bX; \tilde{\bbeta}( \cdot ), \mu_0(\cdot) \big] \Big\}   \\
    & = & \left[ \Pr ( \gamma < \exp\{ \bX^{\T} \tilde{\bbeta}(\tau) \} ) - \Pr ( \gamma < \exp\{ \bX^{\T} \bbeta_0(\tau) \} ) \right]^2  \\
    & = & \left[ E \left[ g( \exp\{ \bX^{\T} \bbeta_*(\tau) \} | \bX ) (  \exp\{ \bX^{\T} \tilde{\bbeta}(\tau) \} -  \exp\{ \bX^{\T} \bbeta_0(\tau) \}  ) \right]  \right]^2,
\end{eqnarray*}
where $\bbeta_*(\tau)$ is between $\tilde{\bbeta}(\tau)$ and $\bbeta_0(\tau)$.
Then $ \mbox{II}_2 = o_p(1) $ follows immediately from the boundedness property of the density function $g$, the continuity of $\exp(x)$, and the uniform consistency of $\tilde{\bbeta}(\tau)$.
For $\mbox{II}_3$, simple algebra implies that
\begin{eqnarray*}
 \mbox{II}_3  & \leq & 2 \cdot E \left\{  \int_{\gamma} \left[ f\{ \gamma | m, C, \bX; \tilde{\bbeta}( \cdot ), \mu_0(\cdot) \}  -  f\{ \gamma | m, C, \bX; \bbeta_0( \cdot ), \mu_0(\cdot) \} \right]^2   d \gamma \right\}   \\
 & \leq & 4 \cdot E \left\{  \int_{\gamma} \left[ f\{ \gamma | m, C, \bX; \tilde{\bbeta}( \cdot ), \mu_0(\cdot) \}  -  f\{ \gamma | m, C, \bX; \check{\bbeta}_0( \cdot ), \mu_0(\cdot) \} \right]^2 d \gamma \right\}  \\
 & &  + 4 \cdot E \Big\{  \int_{\gamma} \left[ f\{ \gamma | m, C, \bX; \check{\bbeta}_0( \cdot ), \mu_0(\cdot) \}  -  f\{ \gamma | m, C, \bX; \bbeta_0( \cdot ), \mu_0(\cdot) \} \right]^2  \\
 & &   ~~~~~~~~~~~~~  I\{ \bX^{\T} \bbeta_0(\tau_1) < \log(\gamma) \leq \bX^{\T} \bbeta_0(\tau_K)   \}  d \gamma \Big\}   \\
 & &  + 4 \cdot E \Big\{  \int_{\gamma} \left[ f\{ \gamma | m, C, \bX; \check{\bbeta}_0( \cdot ), \mu_0(\cdot) \}  -  f\{ \gamma | m, C, \bX; \bbeta_0( \cdot ), \mu_0(\cdot) \} \right]^2  \\
 & &   ~~~~~~~~~~~~~        I\{ \log(\gamma) \leq  \bX^{\T} \bbeta_0(\tau_1)  \}  d \gamma \Big\}   \\
 & &  + 4 \cdot E \Big\{  \int_{\gamma} \left[ f\{ \gamma | m, C, \bX; \check{\bbeta}_0( \cdot ), \mu_0(\cdot) \}  -  f\{ \gamma | m, C, \bX; \bbeta_0( \cdot ), \mu_0(\cdot) \} \right]^2   \\
 & &   ~~~~~~~~~~~~~        I\{ \log(\gamma) > \bX^{\T} \bbeta_0(\tau_K)  \} d \gamma \Big\}   \\
 & \doteq & \mbox{Q}_1 + \mbox{Q}_2 + \mbox{Q}_3 + \mbox{Q}_4,
\end{eqnarray*}
where $ \check{\bbeta}_0( \cdot ) \in \mathcal{G}$ is the right continuous piecewise linear function which satisfy $\check{\bbeta}_0(\tau_k)=\bbeta_0(\tau_k)$.
It follows from Lemma \ref{lem1} that $\mbox{Q}_1 = o(1)$, and from Lemma \ref{lem3} that $\mbox{Q}_2 = o(1)$.
 { By} the assumptions $\Pr( \gamma \leq \exp\{\bX^{\T} \bbeta_0(\tau_1)\} ) = o(1)$, $\Pr( \gamma > \exp\{\bX^{\T} \bbeta_0(\tau_K)\} ) = o(1)$, and the boundedness of $f\{ \gamma | m, C, \bX; \check{\bbeta}( \cdot ), \mu_0(\cdot) \}$ and $f\{ \gamma | m, C, \bX; \bbeta_0( \cdot ), \mu_0(\cdot) \}$ implied by condition C1 and C4(a), we have $\mbox{Q}_3 = o_p(1)$ and $\mbox{Q}_4=o_p(1)$.
Hence we complete the proof of Lemma \ref{lem4}.

{\it Proof of Theorem \ref{theo2}.}
First, we show that $n^{1/2} \bS_n( \hat{\bbeta}, \hat{\mu}, \tau ) = o(1)$, a.s. under the assumption $n^{1/2} \| \mathcal{S}_K \| \rightarrow 0$.
By the definition of $\bS_n(\hat{\bbeta}, \hat{\mu}, \tau)$ and the regularity condition C1, their exists a constant $C_4$ such that
\begin{eqnarray*}
 & &  \sup_{ \tau \in [\tau_k, \tau_{k+1}] } n^{1/2} \| \bS_n( \hat{\bbeta}, \hat{\mu}, \tau ) - \bS_n( \hat{\bbeta}, \hat{\mu}, \tau_k )   \|  \leq n^{1/2} C_4 \cdot | \tau - \tau_k |    \\
 & \leq & C_4 \cdot n^{1/2} \| \mathcal{S}_K \| = o(1), ~\mbox{a.s.}
\end{eqnarray*}
According to the uniform consistency of $\hat{\bbeta}(\cdot)$ by Theorem \ref{theo1}, the uniform consistency of $\hat{\mu}(\cdot)$ by Lemma \ref{lem0}, Lemma \ref{lem4}, and the fact $n^{1/2} \bS_n( \hat{\bbeta}, \hat{\mu}, \tau ) = o(1)$, we have
\begin{eqnarray*}
& & - n^{1/2} \bS_n( \bbeta_0, \mu_0, \tau )     \\
&=& n^{1/2} \{ \bs ( \hat{\bbeta}, \hat{\mu}, \tau ) - \bs( \bbeta_0, \mu_0, \tau ) \} + o_{[\tau_1, \tau_K]}(1)    \\
&=& n^{1/2} \{ \bs ( \hat{\bbeta}, \hat{\mu}, \tau ) - \bs ( \hat{\bbeta}, \mu_0, \tau ) \} + n^{1/2} \{ \bs ( \hat{\bbeta}, \mu_0, \tau ) - \bs( \bbeta_0, \mu_0, \tau ) \} + o_{[\tau_1, \tau_K]}(1)  \\
&=& n^{1/2} \{ \bs ( \bbeta_0, \hat{\mu}, \tau ) - \bs ( \bbeta_0, \mu_0, \tau ) \} + n^{1/2} \{ \bs ( \hat{\bbeta}, \mu_0, \tau ) - \bs( \bbeta_0, \mu_0, \tau ) \} + o_{[\tau_1, \tau_K]}(1),
\end{eqnarray*}
where the last equality holds because of the uniform consistency of $\hat{\bbeta}(\cdot)$, the boundedness of $\bX$, Lemmas \ref{lem1} and \ref{lem3}.
Hence
\begin{eqnarray}
 & & n^{1/2} \{ \bupsilon( \hat{\bbeta}) - \bupsilon(\bbeta_0) \}  \nonumber     \\
 &=& n^{1/2} \{ \bs ( \hat{\bbeta}, \mu_0, \tau ) - \bs( \bbeta_0, \mu_0, \tau ) \}  \nonumber  \\
 &=& - n^{1/2} \bS_n( \bbeta_0, \mu_0, \tau )  - n^{1/2} \{ \bs ( \bbeta_0, \hat{\mu}, \tau ) - \bs ( \bbeta_0, \mu_0, \tau ) \}  + o_{[\tau_1, \tau_K]}(1)  \nonumber  \\
 &=& - n^{-1/2} \sum_{i=1}^n \boldsymbol{\eta}_{1i}(\tau)  - n^{1/2} \{ \bs ( \bbeta_0, \hat{\mu}, \tau ) - \bs ( \bbeta_0, \mu_0, \tau ) \}  + o_{[\tau_1, \tau_K]}(1), \qquad
\end{eqnarray}
where $\boldsymbol{\eta}_{1i}(\tau) = \int_{\gamma} \psi_{\tau} \{ \log(\gamma) - \bX_i^{\T} \bbeta_0(\tau) \} \cdot \bX_i \cdot f\{ \gamma |m_i, C_i, \bX_i; \bbeta_0(\cdot), \mu_0(\cdot) \} d \gamma$.
To approximate $n^{1/2} \{ \bs ( \bbeta_0, \hat{\mu}, \tau ) - \bs ( \bbeta_0, \mu_0, \tau ) \}$, for all $\bbeta \in \mathcal{G}$, we { use} the functional derivative of $\bs(\bbeta, \mu, \tau)$ at $\mu_0$ in the direction $[\mu - \mu_0]${, which is given by}
\begin{eqnarray*}
& & \Gamma_2(\bbeta, \mu_0, \tau)[\mu-\mu_0]   \\
&=& \lim_{ \epsilon \rightarrow 0 } \frac{1}{\epsilon} \big[ \bs\{ \bbeta, \mu_0 + \epsilon(\mu-\mu_0), \tau \} - \bs( \bbeta, \mu_0, \tau ) \big]     \\
&=& \lim_{ \epsilon \rightarrow 0 } \frac{1}{\epsilon} E \big[ \int_{\gamma} \bX \cdot \psi_{\tau} \{ \log(\gamma) - \bX^{\T}\bbeta \}    \\
& & ~~~~~~~~~~ \cdot \{ f\{ \gamma|m, C, \bX; \bbeta(\cdot), \mu_0+\epsilon(\mu-\mu_0) \} - f\{ \gamma|m, C, \bX; \bbeta(\cdot), \mu_0 \} \} d \gamma  \big]    \\
&=&  E \big[ \int_{\gamma} \bX \cdot \psi_{\tau} \{ \log(\gamma) - \bX^{\T}\bbeta \}    \\
& & ~~~~~~~~~  \lim_{ \epsilon \rightarrow 0 } \frac{1}{\epsilon} \{ f\{ \gamma|m, C, \bX; \bbeta(\cdot), \mu_0+\epsilon(\mu-\mu_0) \} - f\{ \gamma|m, C, \bX; \bbeta(\cdot), \mu_0 \} \} d \gamma  \big]    \\
&=&  E \Big[ \int_{\gamma} \bX \cdot \psi_{\tau} \{ \log(\gamma) - \bX^{\T}\bbeta \}  \big\{ \frac{ \phi(m, \gamma, C; \mu_0)[\mu - \mu_0] g\{ \gamma|\bX; \bbeta(\cdot) \}  }{  \int_{\gamma} \rho\{ m|\gamma, C; \mu_0(\cdot) \} g\{ \gamma|\bX; \bbeta(\cdot) \}  d \gamma }  \\
& & ~~~  - \frac{ \rho\{ m|\gamma, C; \mu_0(\cdot) \} g\{ \gamma|\bX; \bbeta(\cdot) \} \int_{\gamma} \phi(m, \gamma, C; \mu_0)[\mu - \mu_0] g\{ \gamma|\bX; \bbeta(\cdot) \}  d \gamma }{ \{ \int_{\gamma} \rho\{ m|\gamma, C; \mu_0(\cdot) \} g\{ \gamma|\bX; \bbeta(\cdot) \}  d \gamma \}^2 }  \big\}  \Big],
\end{eqnarray*}
where
\begin{eqnarray*}
& & \phi(m, \gamma, C; \mu_0)[\mu - \mu_0] \\
& \doteq & \lim_{ \epsilon \rightarrow 0 } \frac{1}{\epsilon} \big[ \rho\{m|\gamma, C; \mu_0 + \epsilon(\mu-\mu_0) \} - \rho\{m|\gamma, C; \mu_0 \} \big]   \\
&=& \lim_{ \epsilon \rightarrow 0 } \frac{1}{\epsilon} \big[ \frac{ \gamma^m \{ \mu_0(C) + \epsilon ( \mu(C) - \mu_0(C) ) \}^m }{ m! } \exp\{ -\gamma \{ \mu_0(C) + \epsilon ( \mu(C) - \mu_0(C) ) \} \}  \\
& & ~~~~~ - \frac{ \gamma^m \{ \mu_0(C) \}^m }{ m! } \exp\{ -\gamma \{ \mu_0(C) + \epsilon ( \mu(C) - \mu_0(C) ) \} \} \big]   \\
& & + \lim_{ \epsilon \rightarrow 0 } \frac{1}{\epsilon} \big[ \frac{ \gamma^m \{ \mu_0(C) \}^m }{ m! } \exp\{ -\gamma \{ \mu_0(C) + \epsilon ( \mu(C) - \mu_0(C) ) \} \}  \\
& & ~~~~~ - \frac{ \gamma^m \{ \mu_0(C) \}^m }{ m! } \exp\{ -\gamma \mu_0(C) \} \big]    \\
&=& \frac{ m \gamma^m \{\mu_0(C)\}^{m-1} \{ \mu(C) - \mu_0(C) \} }{ m! } \exp\{ -\gamma \mu_0(C) \} \\
& & + \frac{ \gamma^m \{ \mu_0(C)^m \} }{ m! } \exp\{ -\gamma \mu_0(C) \} \{ -\gamma \{ \mu(C) - \mu_0(C)  \} \}  \\
&=& \frac{ \gamma^m \{ \mu_0(C) \}^{m-1} \exp\{ -\gamma \mu_0(C) \}  }{ m! } \{ m-\gamma \mu_0(C) \} \{ \mu(C) - \mu_0(C) \}   \\
& \doteq & \kappa(m, \gamma, C; \mu_0) \{ \mu(C) - \mu_0(C) \},
\end{eqnarray*}
with $\kappa(m, \gamma, C; \mu_0) = \gamma^m \{ \mu_0(C) \}^{m-1} \exp\{ -\gamma \mu_0(C) \} \{ m-\gamma \mu_0(C) \}/ m!$. Thus
\begin{eqnarray*}
& & n^{1/2} \{ \bs ( \bbeta_0, \hat{\mu}, \tau ) - \bs ( \bbeta_0, \mu_0, \tau ) \} = n^{1/2}  \Gamma_2(\bbeta_0, \mu_0, \tau)[\hat{\mu}-\mu_0] + o_{[\tau_1, \tau_K]}(1)  \\
&=& E \Big[ \int_{\gamma} \bX \cdot \psi_{\tau} \{ \log(\gamma) - \bX^{\T}\bbeta_0(\tau) \}  \big\{ \frac{  \kappa(m, \gamma, C; \mu_0) g\{ \gamma|\bX; \bbeta_0(\cdot) \}  }{  \int_{\gamma} \rho\{ m|\gamma, C; \mu_0(\cdot) \} g\{ \gamma|\bX; \bbeta_0(\cdot) \}  d \gamma }  \\
& & ~~~  - \frac{ \rho\{ m|\gamma, C; \mu_0(\cdot) \} g\{ \gamma|\bX; \bbeta_0(\cdot) \} \int_{\gamma}  \kappa(m, \gamma, C; \mu_0) g\{ \gamma|\bX; \bbeta_0(\cdot) \}  d \gamma }{ \{ \int_{\gamma} \rho\{ m|\gamma, C; \mu_0(\cdot) \} g\{ \gamma|\bX; \bbeta_0(\cdot) \}  d \gamma \}^2 }  \big\}  \Big]   \\
& &~~~~~ \cdot \{ \hat{\mu}(C) - \mu_0(C) \} + o_{[\tau_1, \tau_K]}(1)   \\
&\doteq&  n^{-1/2} \sum_{i=1}^n \boldsymbol{\eta}_{2i}(\tau) + o_{[\tau_1, \tau_K]}(1),
\end{eqnarray*}
where
\begin{eqnarray*}
\boldsymbol{\eta}_{2i}(\tau) &=& E \Big[ \int_{\gamma} \bX \cdot \psi_{\tau} \{ \log(\gamma) - \bX^{\T}\bbeta_0(\tau) \}  \big\{ \frac{  \kappa(m, \gamma, C; \mu_0) g\{ \gamma|\bX; \bbeta_0(\cdot) \}  }{  \int_{\gamma} \rho\{ m|\gamma, C; \mu_0(\cdot) \} g\{ \gamma|\bX; \bbeta_0(\cdot) \}  d \gamma }  \\
& & ~~~  - \frac{ \rho\{ m|\gamma, C; \mu_0(\cdot) \} g\{ \gamma|\bX; \bbeta_0(\cdot) \} \int_{\gamma}  \kappa(m, \gamma, C; \mu_0) g\{ \gamma|\bX; \bbeta_0(\cdot) \}  d \gamma }{ \{ \int_{\gamma} \rho\{ m|\gamma, C; \mu_0(\cdot) \} g\{ \gamma|\bX; \bbeta_0(\cdot) \}  d \gamma \}^2 }  \big\}  \Big]  \\
& & ~~~ \cdot  \varphi_i(C) \mu_0(C),
\end{eqnarray*}
and the last equality holds from Corollary \ref{cor1}.

{Secondly}, we show that $\{ \boldsymbol{\eta}_{1i}(\tau) + \boldsymbol{\eta}_{2i}(\tau), \tau \in [\tau_1, \tau_K] \}$ is a Donsker class. This follows by the fact that $\{ \psi_{\tau} \{ \log(\gamma) - \bX^{\T} \bbeta_0(\tau) \}, \tau \in [\tau_1, \tau_K]  \}$ is a Donsker class \citep{van1996}, the boundedness properties of $\bX$, $\rho\{ m|\gamma, C; \mu_0(\cdot) \}$, $g\{ \gamma|\bX; \bbeta_0(\cdot) \}$, $\kappa(m, \gamma, C; \mu_0)$ and $f\{ \gamma|m, C, \bX; \bbeta_0(\cdot), \mu_0(\cdot) \}$, along with the permanence properties of the Donsker class. By the Donsker theorem, $n^{1/2} \{ \bupsilon( \hat{\bbeta}) - \bupsilon(\bbeta_0) \} $ converges weakly to a tight Gaussian process $\bG(\tau)$ with mean {\bf 0} and covariance $\bSigma(s, t)$ for $s, t \in [\tau_1, \tau_K]$, where $\bSigma(s, t) = E [ \boldsymbol{\eta}_i(s) \boldsymbol{\eta}_i(t)^{\T} ]$ with $\boldsymbol{\eta}_i(\tau) = \boldsymbol{\eta}_{1i}(\tau) + \boldsymbol{\eta}_{2i}(\tau)$.

Finally, under the regularity condition C3, $\bupsilon(\bbeta)$ is Fr$\acute{\mbox{e}}$chet differentiable at $\bbeta_0(\cdot)$ with continuously invertible derivative $\dot{\bupsilon}_{\bbeta_0}$.
Therefore, $\sqrt{n} \{ \hat{\bbeta}(\tau) - \bbeta_0(\tau) \}$ converges weakly to $ \dot{\bupsilon}_{\bbeta_0}^{-1} \{ \bG(\tau) \}$.

\bibliographystyle{apalike}      
\bibliography{recurrent}   

%
%
\end{document}